\documentclass[10pt,journal,cspaper,compsoc]{IEEEtran}

\ifCLASSOPTIONcompsoc
\else
 \usepackage{cite}
\fi

\ifCLASSINFOpdf
\else
\fi

\ifCLASSOPTIONcompsoc
  \usepackage[caption=false,font=normalsize,labelfont=sf,textfont=sf]{subfig}
\else
  \usepackage[caption=false,font=footnotesize]{subfig}
\fi

\usepackage{graphicx}
\usepackage{amsmath}
\usepackage{amssymb}
\usepackage{color}
\usepackage{latexsym}
\usepackage{epsfig}
\usepackage{amsfonts}
\usepackage{upgreek}
\usepackage{cite}
\usepackage{url}
\usepackage{hyperref}
\usepackage{comment}
\usepackage{multirow} 

\usepackage{rotating}

\usepackage{paralist}

\newcommand{\lie}{\textsf{lie}}
\newcommand{\stress}{\textsf{stress}}

\newcommand{\groups}{\textsf{groups}}
\newcommand{\barycenter}{\textsf{barycenter}}

\newtheorem{Hypothesis}{Hypothesis}
\def\true{\mathop{\textsf{true}}\nolimits}
\def\false{\mathop\textsf{false}}

\begin{document}

\title{Towards Faithful Graph Visualizations}

\author{Quan~Hoang~Nguyen,
        Peter~Eades
\IEEEcompsocitemizethanks{\IEEEcompsocthanksitem Q. Nguyen and P. Eades are with the School of Information Technologies, University of Sydney, Australia.\protect\\
E-mail: \{quan.nguyen,peter.eades\}@sydney.edu.au
}
\thanks{}}

\markboth{Journal of \LaTeX\ Class Files,~Vol.~1, No.~1, January~2016}%
{Shell \MakeLowercase{\textit{et al.}}: Towards Faithful Graph Visualizations}


\IEEEcompsoctitleabstractindextext{%
\begin{abstract}
Readability criteria have been studied as a measurement of the quality of graph visualizations. This paper argues that readability criteria are necessary but not sufficient. We propose a new kind of criteria, namely faithfulness, to evaluate the quality of graph layouts. Specifically, we introduce a general model for quantifying faithfulness, and contrast it with the well established readability criteria. We show examples of common visualization techniques, such as multidimensional scaling, edge bundling and several other visualization metaphors for the study of faithfulness.
\end{abstract}

}

\maketitle

\IEEEdisplaynotcompsoctitleabstractindextext

\IEEEpeerreviewmaketitle


\ifCLASSOPTIONcompsoc
  \noindent\raisebox{2\baselineskip}[0pt][0pt]%
  {\parbox{\columnwidth}{\section{Introduction}\label{sec:introduction}%
  \global\everypar=\everypar}}%
  \vspace{-1\baselineskip}\vspace{-\parskip}\par
\else
  \section{Introduction}\label{sec:introduction}\par
\fi

\IEEEPARstart{V}{isualization} aids visual acuities for pattern detection and knowledge discovery by enabling transformation or deduction from information to knowledge through visual means.
Visualization has been widely used for illustrating, formulating hypothesis, identifying patterns, constructing knowledge, performing tasks and making decisions.

Technological advances have dramatically increased the volume of network data. Graphs generated in modern applications become larger and more complex. Since then, graph visualization techniques are getting more sophisticated and involving complex parameter settings to turn graphs into drawings. Thus, it is vital to justify how reliable visualizations methods and models are.

``Readable'' pictures of graphs are the common aim that has been addressed in graph drawing algorithms developed over the past 35 years. Here ``readability'' is
measured by {\em aesthetic criteria}, such as:
\begin{compactenum}
\item \emph{Crossings}: the picture should have few edge crossings.
\item \emph{Bends}: the picture should have few edge bends.
\item \emph{Area}: the area of a grid drawing should be small.
\end{compactenum}

Readability criteria have been extensively studied in numerous visualization systems~\cite{di,tamassia2013handbook}.  
Algorithms that attempt to optimise aesthetic criteria have been
successfully embedded in systems for analysis in a wide variety of
domains, from the finance industry to biotechnology.

In this paper, we argue that readability criteria for visualizing
graphs, though necessary, are not sufficient for effective graph
visualization. 

Traditionally, it is quite commonplace for the ``presumption'' that quality of visualization is measured by the readability of the visualization; see~\cite{di,tamassia2013handbook,purchase1996validating,purchase1997aesthetic}. Such readability criteria presumes that readability implies that the picture is a faithful representation of the data. 
For the last couple of decades, traditional ways of visualization of the graphs use node-link diagrams to visualize the whole graphs, which are comprised of a fairly limited number of nodes and edges. 
Hence, the presumption may be clearly true in these traditional approaches to graph visualization.

However, for modern visualization metaphors, such as 2.5D visualizations, map-based visualizations, matrix representations and their combinations, the readability criteria is not sufficient.  Further, the presumption can be demonstrably false because of the extensive use of clutter reduction techniques, such as edge bundling. For large-scale graphs in modern applications, visualization of the whole graph is not always feasible. Since data reduction and aggregation become commonplace, readability criteria is not sufficient.

\begin{figure}\centering
\subfloat[Original graph]{
\label{fig:concen_orig}\includegraphics[width=.24\textwidth]{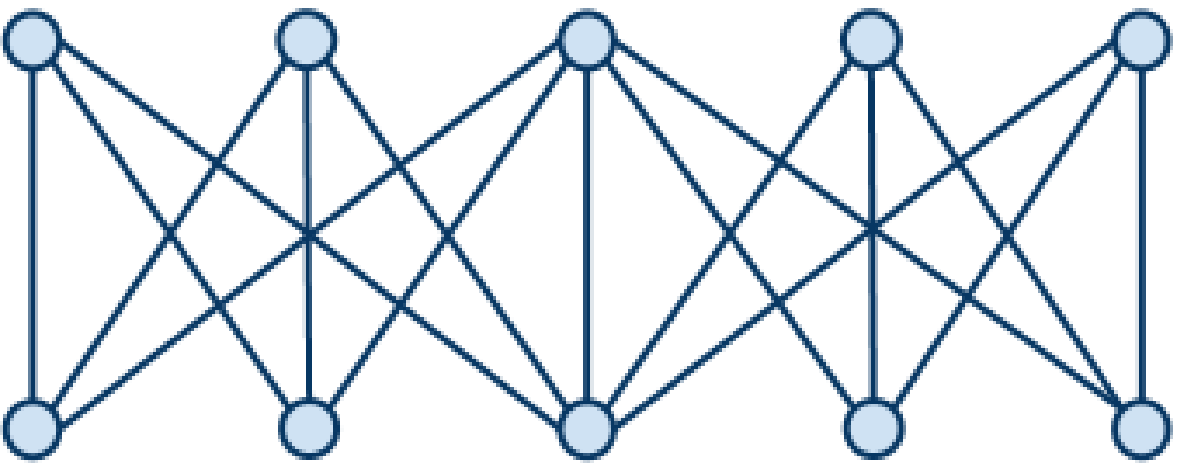}}
\subfloat[Edge contraction]{
\label{fig:origconcentration}\includegraphics[width=.24\textwidth]{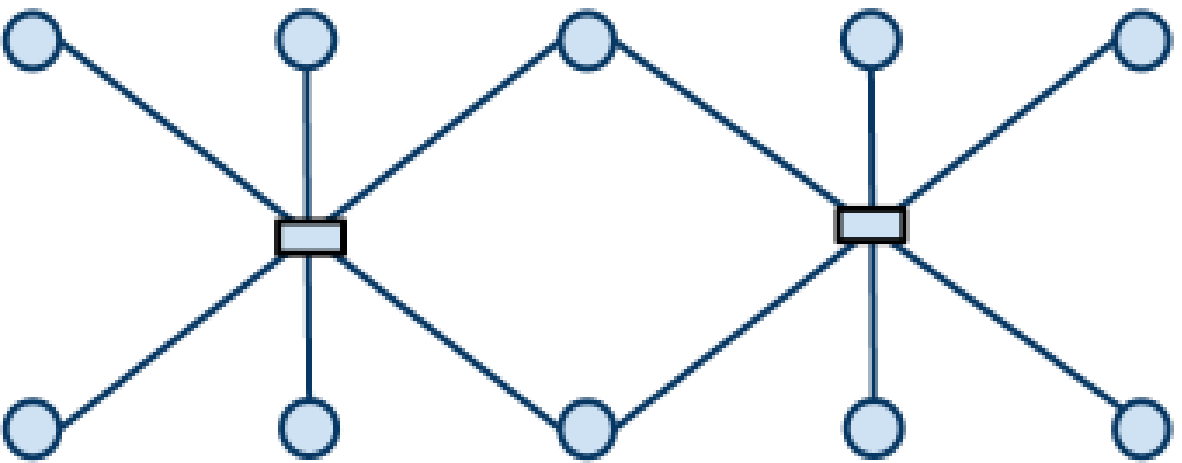}}
\caption{An example of edge concentration~\cite{newbery1989edge}}\label{fig:concentration}
\end{figure}
\begin{figure}\centering
\subfloat[Original graph]{
\label{fig:confluent_orig1}\includegraphics[width=.24\textwidth]{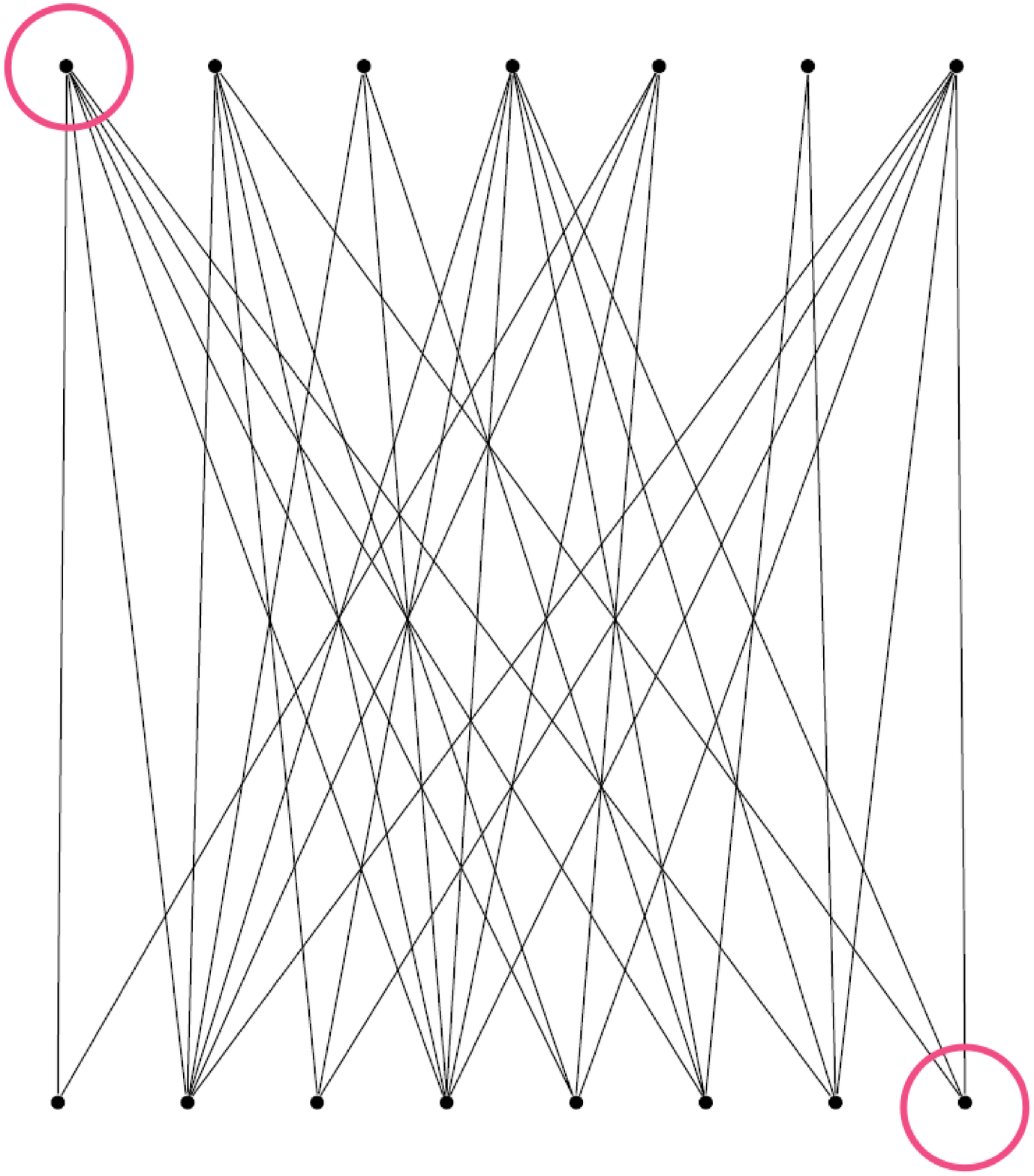}}
\subfloat[Confluent drawing]{
\label{fig:origconfluent}\includegraphics[width=.24\textwidth]{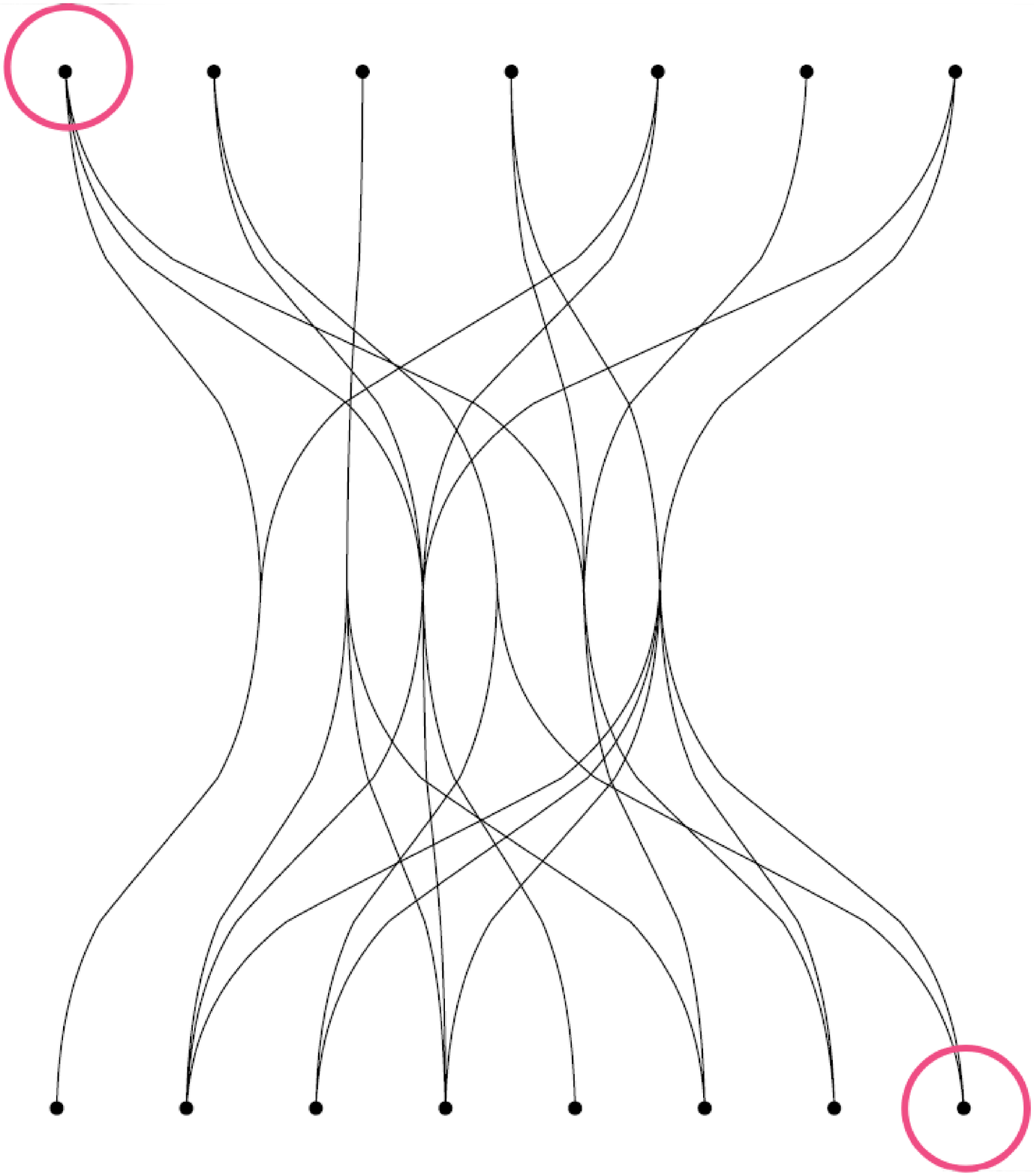}}
\caption{An example of confluent drawing ~\cite{eppstein2007confluent}}\label{fig:confluent}
\end{figure}
Here, we introduce another kind of criterion, generically
called ``faithfulness'', that we believe is necessary in addition to
readability. Intuitively, a graph drawing algorithm is ``faithful'' if
it maps different graphs to distinct drawings\footnote{In Mathematics,
a \emph{faithful representation} of a group on a vector space is a
linear representation in which different elements in the group are
represented by distinct linear mappings.}. Using mathematical terms, a faithful graph
drawing algorithm encodes an \emph{injective} function. In other words, a faithful graph drawing algorithm never maps distinct graphs to the same drawing.

We show several examples below for illustrating the new faithfulness criteria.

\subsection{Motivating examples\label{sec:vizmotivation}}
Many graph visualization algorithms have embedded one or more aesthetic criteria to achieve readable layouts. In some cases however, graph layout algorithms cannot avoid visual clutter or edge cluttering due to high edge density from intrinsic connectivity and overlapping between nodes and edges.

In such cases, \emph{edge concentration}~\cite{newbery1989edge},  \emph{confluent drawing}~\cite{dickerson2004confluent,eppstein2006delta} and \emph{edge bundling}~\cite{holten2006hierarchical,zhou2008energy,gansner2011multilevel}
become useful. These edge routing algorithms share the same idea: they simplify edge connections in the picture to increase readability. Such improved readability is helpful with respect to a number of tasks, yet the pictures may become less faithful.

As an example, \emph{edge concentration}~\cite{newbery1989edge}
simplifies edge connections in the picture to increase readability.
Figure~\ref{fig:concentration} shows two pictures of a bipartite graph;
Figure~\ref{fig:concen_orig} is a simple drawing with straight-line edges
and Figure~\ref{fig:origconcentration} is another drawing with ``concentrated'' edges. Figure~\ref{fig:origconcentration} has less edge crossings and is more readable. However, Figure~\ref{fig:origconcentration} would also give a viewer, who does not know how the picture was made, at the first sight: a graph comprising of ten circle nodes and two boxes connected by twelve lines. This demonstrates the lack of faithfulness of edge concentration.

For our second example, Figure~\ref{fig:confluent_orig1} shows another bipartite graph; the confluent drawing of the graph is depicted in Figure~\ref{fig:origconfluent}. Confluent drawings may be not faithful. For example, one may find there is \emph{no} link connecting the two red circles in Figure~\ref{fig:confluent_orig1}. In contrast, one may see from Figure~\ref{fig:origconfluent} a curve connecting the two red circles. This is clearly an inconsistency in the confluently drawn graph.

\begin{figure}\centering
\subfloat[Unbundled]{
\begin{minipage}[b]{0.45\linewidth}\centering%
{\label{fig:orig}\includegraphics[width=\textwidth]{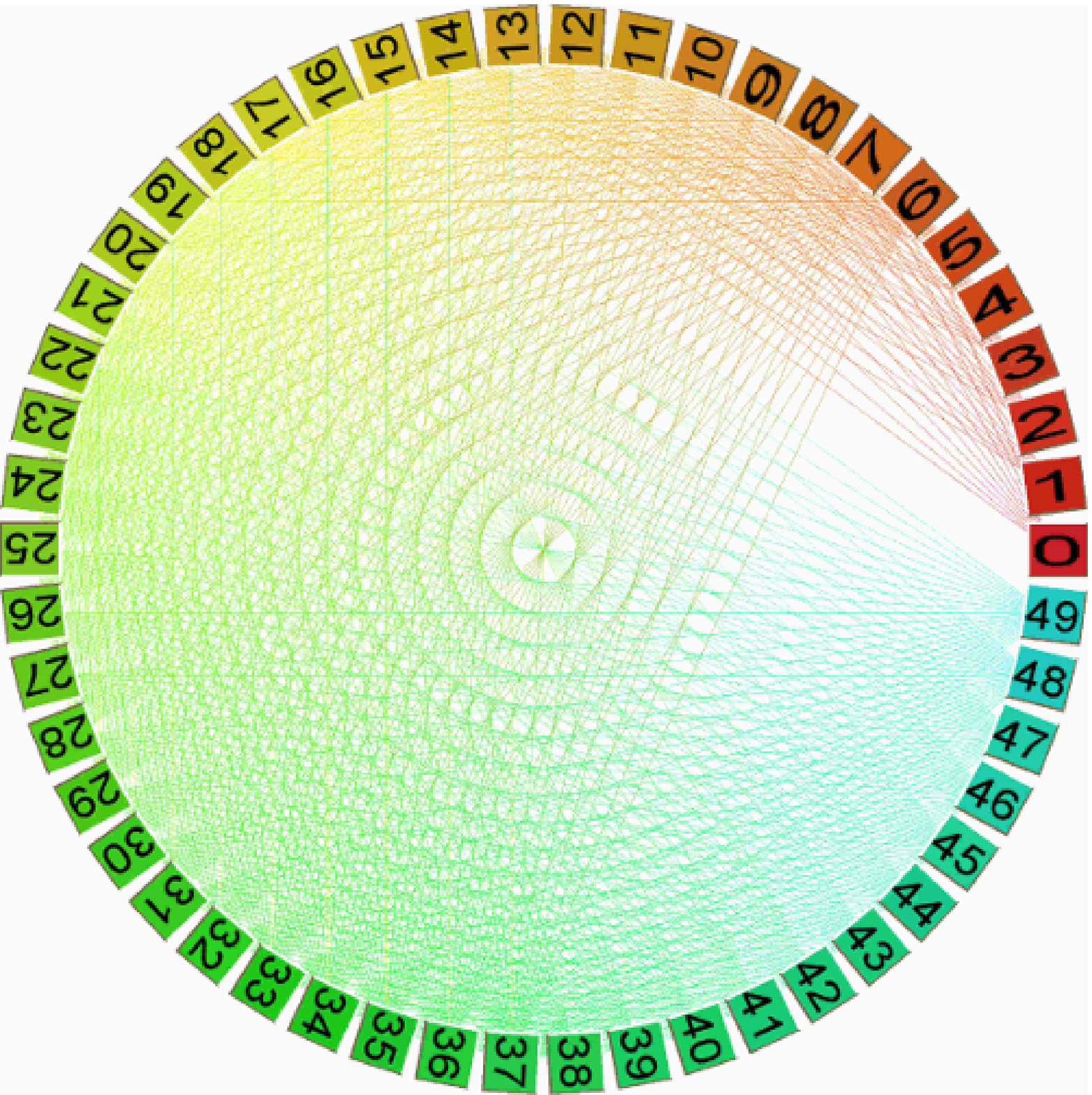}}
\end{minipage}
}\subfloat[Bundled]{
\begin{minipage}[b]{0.45\linewidth}\centering%
{\label{fig:origbundling}\includegraphics[width=\textwidth]{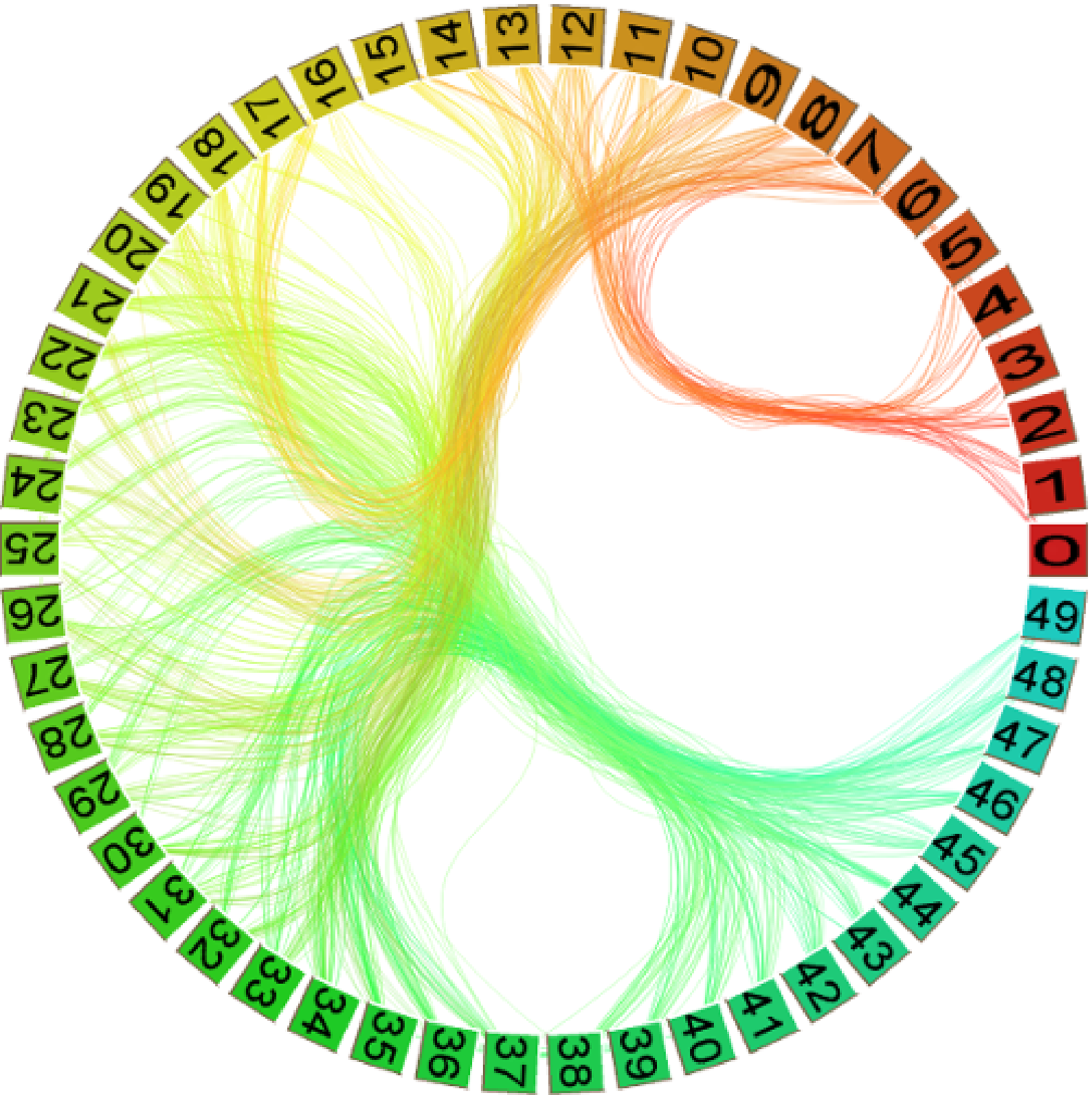}}
\end{minipage}
}
\caption{An example graph using force-directed edge bundling}\label{fig:bundling}
\end{figure}

As an example of edge bundling, Figure~\ref{fig:bundling} shows two pictures of the same graph.
Figure~\ref{fig:orig} is a simple circular layout with straight-line edges
and Figure~\ref{fig:origbundling} has the same node positions but with ``bundled'' edges.

\begin{figure}\centering
\subfloat[Unbundled]{
\begin{minipage}[b]{0.45\linewidth}\centering%
{\label{fig:orig1}\includegraphics[width=\textwidth]{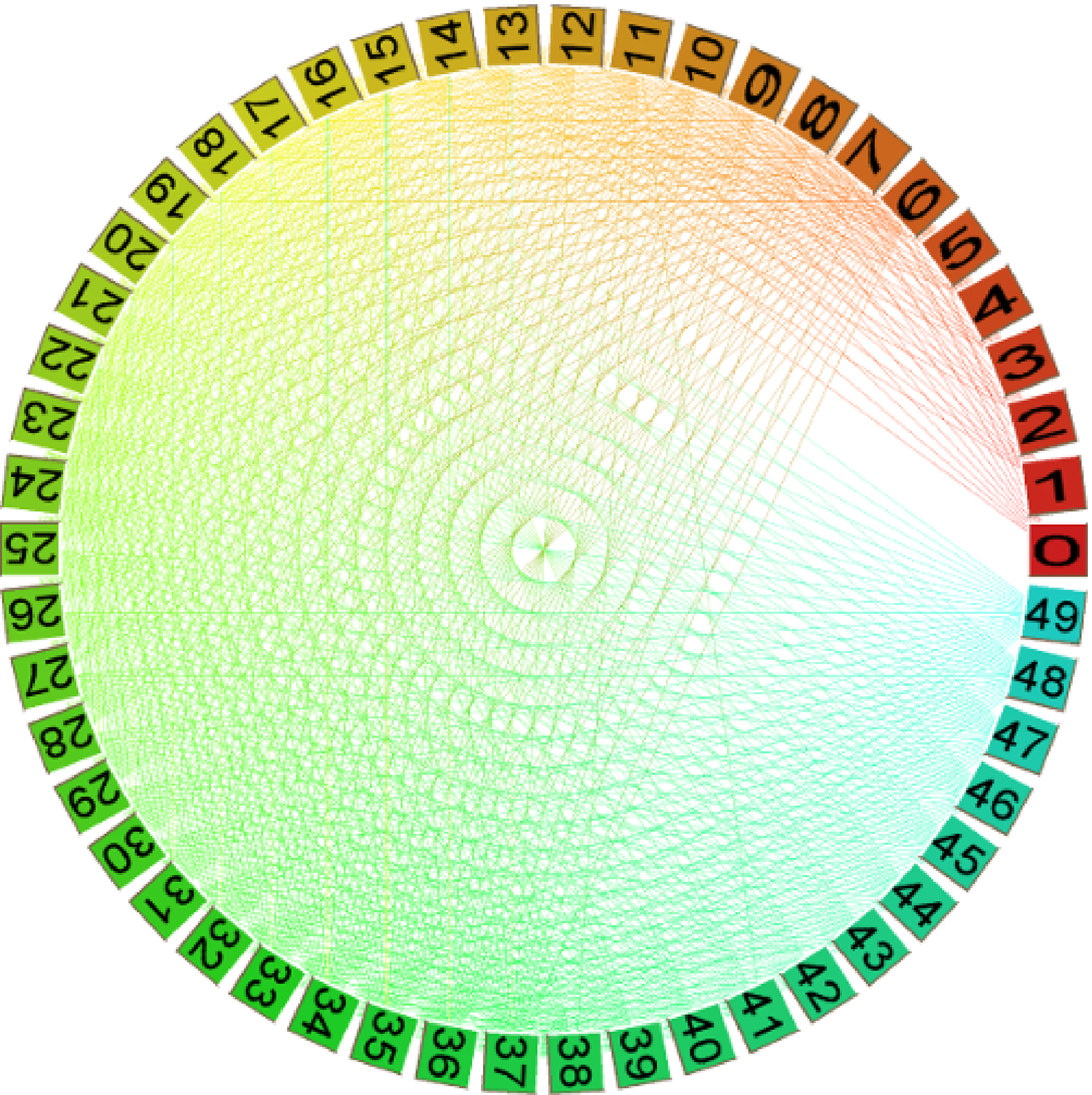}}
\end{minipage}
}
\subfloat[Bundled]{
\begin{minipage}[b]{0.45\linewidth}\centering%
{\label{fig:origbundling1}\includegraphics[width=\textwidth]{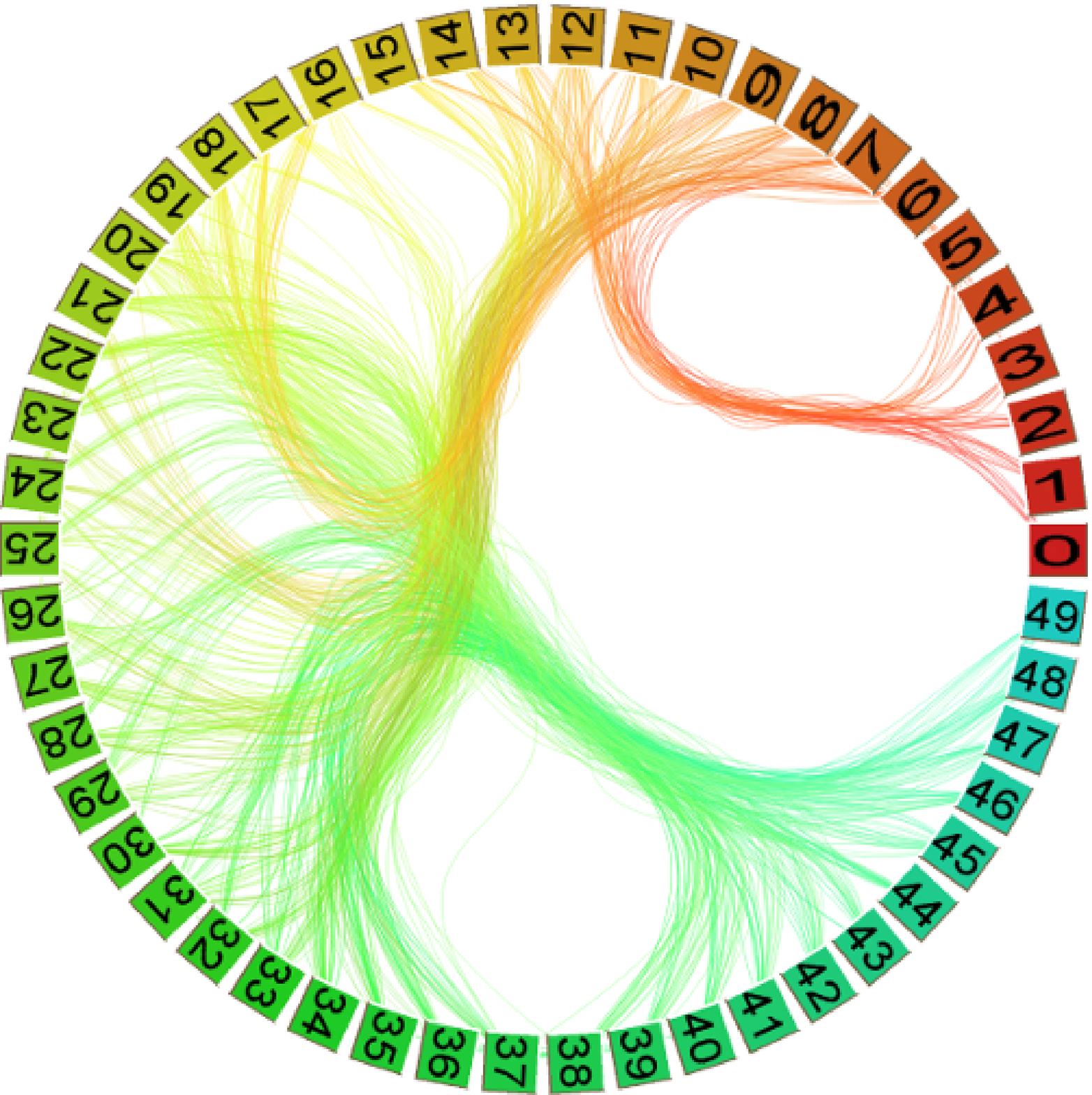}}
\end{minipage}
}
\caption{A 10 percent modification of the example graph in Figure~\ref{fig:bundling}(a) and the result using force-directed edge bundling}\label{fig:bundling1}
\end{figure}

In the example of edge bundling, bundling often sacrifices faithfulness. Remarkably, while the visual effects of bundled edges give a more readable visual representation of the overall graph, the ability to locate, select or navigate individual edges, hopping between nodes is lost. Even worse, bundling can result in situation where two different graphs can be mapped to the same picture. For example, Figure~\ref{fig:orig1} shows a graph that differs from the
graph Figure~\ref{fig:orig} by almost 10 percent of the total number of edges. This graph is bundled in Figure~\ref{fig:origbundling1}.
The bundled representations of the two different graphs (Figure~\ref{fig:origbundling} and 
Figure~\ref{fig:origbundling1}) are identical.

\subsection{Aims and contributions}

The examples in previous section have demonstrated that faithfulness is an important aspect of graph visualizations. 
Yet, faithfulness has yet been paid enough attention by the visualization community. There are a few notions along the lines of faithfulness in scientific visualization, such as {\em fidelity} of the
picture~\cite{matsuyama2004real}, and visual \emph{reconstructability}
for flow visualization~\cite{jänicke2011visual}. 
In addition, several formal models have been proposed~\cite{van2005value,purchase2008theoretical,carpendale2008evaluating,chen2009data,lam2011seven}; they aim for assessing the visualizations as well as for guiding the future of research in Information Visualization.
Nevertheless, there is no model of faithfulness for graph visualization.

In this paper, we distinguish two important concepts: the ``\emph{faithfulness}'' and the \emph{readability} of visualizations of graphs.
Faithfulness criteria are especially relevant for modern methods that
handle very large and complex graphs. Information overload from very large data
sets means that the user can get lost in irrelevant detail, and
methods have been developed to increase readability by decreasing
detail in the picture.

With a vast amount of methods and techniques for visualizing the data sets at various levels of granularity, it is important to know how reliable the visualizations being produced are. This demand has become urgent given the popularity of network data, the information overload from technological advances and various challenges for visualizing large complex and dynamic networks.

In summary, we make the following contributions:
\begin{itemize}
\item{\bf General model of graph visualization:} In order to describe the faithfulness concept, we present a general model of graph visualization process in Section~\ref{sec:vizmodel}.
\item{\bf General model of the faithfulness:} Here, the faithfulness concept is divided into three types of faithfulness: information faithfulness, task faithfulness and change faithfulness.
\item{\bf Examples of faithfulness metrics:} We further define some examples of the faithfulness metrics, which aim to compare the usefulness / effectiveness of different visualizations. We should emphasize that these metrics are just the examples and are not the ``only'' ones.
\item{\bf Case studies of faithfulness:}  As for demonstration, we use popular visualization methods including force-directed methods, multidimensional scaling, matrix representations, and hybrid metaphors to evaluate our faithfulness concept.
\end{itemize}

The rest of the paper is outlined as follows.
Section~\ref{sec:vizmodel} presents our general model of graph visualization process. We use this visualization model to describe the faithfulness concept.
Section~\ref{sec:vizfaithfulness} describes our general model of the faithfulness of a graph visualization method. Here, we divide the general concept into three kinds of faithfulness: information faithfulness, task faithfulness, and change faithfulness.
Section~\ref{sec:vizmetrics} shows our examples of faithfulness metrics, which are computed for quantifying and comparing the faithfulness of different visualizations.
We illustrate faithfulness with several examples in Sections~\ref{sec:vizMDS}, \ref{sec:vizbundling} and \ref{sec:vizmetaphors}: multidimensional scaling, edge bundling and several selected visualization metaphors (matrix-based and
map-based visualizations).
Section~\ref{sec:vizconclusion} concludes the paper with a remark about the failure of 3D graph drawing to make industrial impact.

\section{Related work}

\subsection{Evaluation of visualization}
There have been increasing interests in evaluative research methodologies and empirical work~\cite{van2005value,purchase2008theoretical,carpendale2008evaluating,chen2009data,lam2011seven,lam2012empirical}.

There are several notions along the lines of faithfulness in
scientific visualization. These include {\em fidelity} of the
picture~\cite{matsuyama2004real}, and visual \emph{reconstructability}
for flow visualization~\cite{jänicke2011visual} and {\em faithful} ``functional visualization schema'' introduced by Tsuyoshi et. al.~\cite{Sugibuchi2009}. This notion is to judge the transformation from data schema to visual abstraction schema.

A number of quality metrics have been recently proposed for evaluating high-dimensional data visualization \cite{bertini2011quality}.
A survey of quality concerns for parallel coordinates is given in \cite{dasgupta2011adaptive}.
There is also proposal for judging visualizations regarding the presence of difficulties and distracting elements (or so-called ``chartjunk'') \cite{hullman2011benefitting}.
Other research focuses on narrative visualization and the effects on interpretations with respect to the intended story~\cite{hullman2011visualization}. 

These previous research on visualization evaluation has mainly focused on information visualization in a broad sense, while our research addresses the quality of graph visualizations.

Recently, an algebraic process to evaluate visualization design has been proposed~\cite{kindlmann2014}. Their notions of  \emph{representation invariance}, \emph{unambiguous data depiction} and \emph{visual-data correspondence} are close to our concept of information faithfulness and change faithfulness. They evaluated their approach on bar charts and scatterplots. 
In contrast, our approach aims for the quality (the faithfulness) of graph visualizations.

\subsection{Readability in graph visualization}
Graph drawing algorithms in the past 30 years typically take into account one or more {\em aesthetic criteria} to aim to increase the \emph{readability} of the drawing and to achieve ``nice'' drawings.
There are a wide range of aesthetic criteria proposed for graph visualizations. They include, for example,
(1) minimizing the number of edge crossings minimization~\cite{reingold1981tidier};
(2) minimizing the number of bends in polyline edges~\cite{reingold1981tidier};
(3) increasing orthogonality: placing nodes and edges to an orthogonal grid~\cite{reingold1981tidier};
(4) increasing node distribution~\cite{tamassia1988automatic,davidson1996drawing}.
(5) maximizing minimum edge angles between all edges of a node~\cite{purchase2002metrics};
(6) minimizing the total area~\cite{tamassia1988automatic};
(7) maximizing the symmetries in the underlying network structure~\cite{north};
(8) edge lengths should be short but not too short\cite{coleman1996aesthetics}.

Amongst these aesthetics,  minimizing edge crossings is the most important criterion from previous user studies~\cite{purchase1997aesthetic}.

Optimizing two or more criteria simultaneously is an NP-hard problem in general. As such, graph drawings are often the results of compromise among several aesthetics. Most of previous research has also aimed for computational efficiency while achieving the drawing readability.

In the literature, graph drawing algorithms can be generally classified, for example, in one the following criteria: (1)
A popular technique is force-directed
layout, which uses physical analogies to achieve an
aesthetically pleasing drawing~\cite{kamada1989algorithm,DBLP:books/ph/BattistaETT99,brandes2001drawing}; (2) 
Several works extended force-directed algorithms for drawing large graphs in \cite{harel2001fast, walshaw2001multilevel}; (3)
Multidimensional scaling is another popular method for graph visualization and visual mining~\cite{DBLP:conf/gd/BrandesP06}; (4) Another approach takes advantage of any knowledge on topology (such as planarity or SPQR decomposition~\cite{newbery1989edge,feng1995draw,gutwenger2002graph,mutzel2003spqr,eades2010graph}) to optimize the drawing in terms of readability; (5) Other approaches offer representations
composed of visual abstractions of clusters to improve readability.

The faithfulness criterion proposed in this paper is important to compare the usefulness of graph visualizations. This faithfulness is different from readability criteria in the literature.

\subsection{Mental map preservation in graph visualization}

An important criterion for dynamic graph drawings is {\em mental map preservation}~\cite{peter1991mentalmap} or \emph{stability}\cite{paulisch1990edge}.
The mental map concept refers to the abstract geometric structures of a person's mind while exploring visual information. The better the mental map is preserved, the easier the structural change
of a graph is understood.

To preserve the mental map, some \emph{layout adjustment} algorithms use a notion of proximity
and rearrange a drawing in order to improve some aesthetic criteria. These algorithms include, for example, incremental drawing of directed acyclic graphs~\cite{north1996incremental}, or computing the layout of a sequence of graphs offline~\cite{diehl2002graphs}, or using different adjustment strategies in order to compute the new layout~\cite{huang1998fully}.
Other works have studied the mental map preservation while the graph is being \emph{updated}~\cite{frishman2004dynamic,lee2006mental,lin2011mental}.
This can be achieved by using force directed layout techniques~\cite{frishman2004dynamic}, or
 by using simulated annealing\cite{lee2006mental}.

There are some trade-offs between the readability and the stability of (offline) dynamic graph drawing methods (see a survey by Brandes et al.~\cite{brandes2012quantitative}).

In contrast to the mental map in the previous work, this paper defines ``change faithfulness'', which captures the sensitivity of the pictures to changes in the data. This change faithfulness aims for the understanding and evaluation of dynamic graph visualization.

\subsection{Temporal and spatial analysis}
For analysis of dynamic graphs, it commonly requires to show the statistical trends
and changes over time. Visualization of dynamic graphs also needs to preserve user's mental map~\cite{misue1995layout}.
The most common techniques for representing temporal data are via {\em animation} and {\em the ``small multiples'' display} (see~\cite{archambault2011animation}). The animation approach shows visualizations of the sequence of graphs displayed in consecutive frames. The small multiples display uses multiple charts laid side-by-side and corresponding to consecutive time periods or moments in time~\cite{andrienko2006exploratory}.

Generally speaking, this previous work has been focused on the space dimension. For example,  most of the work considers readability and stability or mental map preservation of 2D / 3D drawings of graphs. In this paper, we are concerned about the faithfulness in both the \emph{space} and the \emph{time} dimensions.

\section{Graph visualization model\label{sec:vizmodel}}

In this section we describe a semi-formal model for
graph visualization. The section first gives basic annotations of graphs and then describes our graph visualization model.

A \emph{graph} $G=(N,E)$ consists a set of nodes $N$ and a set of
edges $E$.  Note that, in this paper we use $N$ to denote the set of nodes and preserve $V$ to denote visualization.
In practice, the nodes and edges may have multiple
attributes, such as textual labels. For example, for Facebook friendship networks, a node may be associated with names, education, marriage status, current position, etc; whereas an edge may have relationship types between two persons. These attributes can be important
for visualization and analysis.
Further, nodes and edges may be timestamped, and the visualization
varies over time.

\begin{figure}\centering
\includegraphics[width=.5\textwidth]{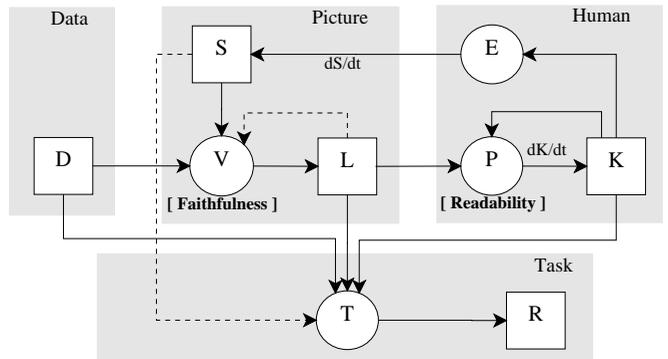}
\caption{Graph visualization model\label{fig:vispipeline}
}
\end{figure}

Figure~\ref{fig:vispipeline} shows our visualization model.
It is an extension of the van Wijk model~\cite{van2005value}.
Our model encapsulates the whole knowledge discovery process, from
data to visualization to human; unlike the van Wijk model, our model includes \emph{tasks}.

The main processes of the model are ``visualization'' $V$, ``perception'' $P$, and ``task'' $T$, and described as follows.

\subsection{Visualization}
The \emph{visualization process} maps 
a \emph{data} item $d \in D$ (an attributed graph) to a \emph{layout} item
(sometimes called a \emph{``picture''}) $\ell = V(d) \in L$ according to
a \emph{specification} $s \in S$. We write this as follows\footnote{For this
semi-formal model, we use mathematical notation more as a concise
shorthand rather than a precise description.
 For example, we describe processes as functions, but we
should warn the reader that the domains and ranges of these are sometimes
ill-defined.}:
\begin{equation}
\label{eq:visualization0}
V : D \times S \rightarrow L,
\end{equation}
with data space $D$, specification space $S$ and layout space $L$

\begin{itemize}
\item The type of data space $D$ to be visualized can vary from a simple list of nodes and edges
to a time-varying graph with complex attributes on nodes and edges. 
There are a vast amount of network data and many of them have been generated in fairly high rates.
\item 
The specification space $S$ includes, for example,
a specification of the hardware used such as the size and the resolution of the screen. The specification can also include parameter inputs for visualization, navigation or interaction.
\item
The layout space $L$ may consist of graph drawings in the usual sense, but more generally consist of structured objects
in a multidimensional geometric space. Sometimes it is convenient to regard the layout space as the screen; in this case, using the language of Computer Graphics, it is an \emph{image space}.
\end{itemize}

In many applications, nodes and edges of a graph in the data space $D$ may contain several attributes. For example, a social network from Facebook have nodes representing people and edges representing their social connections. In this network, nodes (people) may have other information such as age, gender, identity, marital status, education, and many more. Edges may be associated with additional information of whether they are classmate, colleague or family relationships.

Typically, to represent different attributes in the layout space $L$, the drawings may comprise of a variety of visual cues, such as color, shape or transparency. For \emph{classic} drawing algorithms, the visualization process may compute the layout directly from the data.

For \emph{incremental} algorithms, the visualization process may use the
previous layout when computing the current layout.
The capability of modelling incremental algorithms as well as dynamic graphs makes
our visualization model more general than the van Wijk model~\cite{van2005value}.
In this case,
the general form of visualization process becomes:
\begin{equation*}
\label{eq:visualization1}
V : D \times S \times L \rightarrow L ,
\end{equation*}
where the previous layout is the input for computing the current layout.
This is necessary, for example, when a layout algorithm attempts to
preserve the user's mental map. However, unless otherwise stated we take the simpler model in equation (\ref{eq:visualization0}).

\subsection{Perception}
The \emph{perception process} maps a picture from the layout space
$L$ to the \emph{knowledge} space $K$. We write this as
follows:
\begin{equation}
\label{eq:visualization2}
P : L \rightarrow K
\end{equation}
In this model we use the term {\em knowledge} -- sometimes called
\emph{insight} or \emph{mental picture} -- to denote the effect on the
human of his/her observation of the picture. Again, in time-varying
situation, the human's perception can depend on the previous picture,
and perhaps it is better to write:
\begin{equation*}
P : L \times K \rightarrow K
\end{equation*}
However, we use the simpler model (\ref{eq:visualization2}) unless otherwise stated.

Of course, it is difficult to formally model human knowledge or insight, and
it is difficult to assess its value~\cite{north2006toward}. In particular, in some
situations such as \emph{exploratory} visualization, the information
that is contained in the data is not known \emph{a priori}, and we make
pictures to get serendipitous insight.
This is a well-known paradox; we do not know in advance the aspects or features that should be visible and thus it is hard to assess how successful we are.

In terms of perception, the human perceptual system has several constraints. Some visual channels are probably more representational, expressive or perceivable than others. For instance, size and length are more effective for \emph{quantitative} data. Yet, for \emph{ordinal} or \emph{nominal} data, size and length are less useful.
In other example, one person can distinguish some pairs / groups of colors more easily than other persons can. Sometimes, the effectiveness of a visual channel may vary between people; different people may perceive a picture differently.

\subsection{Task}
Visualizations are a useful means for exploration and examination of data using visual representations or pictures. In many practical cases, visualizations are developed to serve for domain-specific \emph{tasks}~\cite{treinish1998task}. 
Examples of common tasks include, for example, identifying important actors and communities in a social network, or exploring possible pathways in a biological network. To understand faithfulness, it is important to model these tasks. 

Generally speaking, tasks can be distinguished as ``low-level'' tasks~\cite{Wehrend1990} and  ``high level'' tasks~\cite{hibino1999task}. 
For example, Wehrend and Lewis~\cite{Wehrend1990} gives a list of possible low-level tasks (e.g., identify, locate, distinguish, categorize, cluster, distribute, rank, compare, associate and correlate) that one can perform for data analysis. In Hibino's study~\cite{hibino1999task} of a data set of tuberculosis, seven high-level tasks include prepare, plan, explore, present, overlay, reorient, and other.

This paper models all tasks as processes that map the data space $D$,
the layout space $L$, and the knowledge space $K$ to a
\emph{result space} $R$. 
Note that, our graph visualization model differs from the van Wijk model~\cite{van2005value} by the task model. The result space $R$ may be a simple boolean
space $\{ \true , \false \}$; or more commonly it is a multidimensional
space, with each dimension modelling a separate
subtask.

Tasks are often executed by users or data analysers. However, tasks are not necessarily performed by the same person, who creates the pictures from the data. 
Furthermore, tasks can be performed directly by extracting the answers from the data with or without using a picture of that data. 

The central model for the task process is $T = (T_D, T_L, T_K)$, where $T_D$, $T_L$ and $T_K$ are three functions:
\begin{align*}
\label{eq:task}
T_D &: D \rightarrow R\\
T_L &: L \rightarrow R\\
T_K &: K \rightarrow R.
\end{align*}

We extend this common notion with two less common notions $T_D$ and $T_L$. These functions are more abstract; they do not take the user’s perception or knowledge into account. The function $T_D$ returns a result directly from the data. Intuitively, one can imagine that $T_D$ is computed by a ``data oracle'', who can extract a result perfectly from the data.

Similarly, one can think of the function $T_L$ as returning a result directly from the picture. Again, one can imagine a ``picture oracle'', who extracts perfect results from the picture. If the visualization mapping creates a picture that is not entirely consistent with the data, then it is possible for the picture oracle to return a different result from the data oracle. In this way, the picture oracle may be limited by the faithfulness of the visualization mapping.

In this model, all the details (such as questions) of a task are considered less important. It is similar to how we do ignore the details (such as algorithms/mechanisms) of a visualization method. Thus, our model is not concerned with specific questions of tasks.

This task model is perhaps over-simplistic; for example, it does not model the \emph{a priori} knowledge of the human. In addition, the task process can be made more general by taking some combination of data, layout and knowledge and then mapping to the result space. However, the simple model is sufficient to demonstrate the concept of faithfulness - task faithfulness. We use this simple model of task throughout the rest of the paper.

\section{Faithfulness model\label{sec:vizfaithfulness}}

Informally, a graph visualization is {\em faithful} if the underlying
network data and the visual representation are logically consistent.
In this section, we develop this intuition into a semi-formal model. In
fact, we distinguish three kinds of faithfulness: \emph{information}
faithfulness, \emph{task} faithfulness, and \emph{change}
faithfulness. Then we discuss the difference between faithfulness and correctness, and then between faithfulness and readability.

\subsection{Information faithfulness}
The simplest form of faithfulness is \emph{information faithfulness}.
This is based on the idea that the visual representation of a data set
should contain all the information of the data set, irrespective of
tasks. In terms of the notation developed above, a visualization $V$
is \emph{information faithful} if the function $V$ is \emph{injective}, that
is, $V$ has an inverse.

As an example, consider the classical $\barycenter$ visualization function that takes as input a planar graph $G=(N,E)$, places nodes from a specified face on the vertices of a convex polygon, and places every other node at the barycenter of its neighbors (see~\cite{tamassia2013handbook}). This function is information faithful on \emph{internally triconnected} (see~\cite{tamassia2013handbook}) planar graphs. However, if the input graph is not internally triconnected, then same picture can result from several input graphs (each internal triconnected components is collapsed onto a line), and the method is not information faithful.

\subsection{Task faithfulness}
The intuition behind \emph{task faithfulness} is that the
visualization should be accurate enough to correctly perform tasks. In
terms of the functions $V$ and $T$ defined above, a visualization $V$ is \emph{task
faithful} with respect to specification $s \in S$ if
\begin{equation}
T_L( V (d, s)) = T_D(d)
\end{equation}
for every data item $d \in D$.

If a visualization is information faithful, then clearly it is task
faithful, assuming the “picture oracle” can extract
all information from the picture to perform tasks. In
many practical cases, such extraction may depend on the perception skills of the viewers.
However, the converse may not hold.

Consider, for example, a visualization function $V_{cir}$ that draws all nodes a graph $G$ on the circle.
Clearly, $V_{cir}$ is information faithful; Figure~\ref{fig:orig} is an example, in which we can find all nodes and edges in the drawing. Further, $V_{cir}$ is task-faithful: all the data is represented in the drawing, and so all tasks can be performed correctly using the drawing.

On the other hand, Figure~\ref{fig:origbundling} shows a graph drawing using edge bundling. Consider the task to estimate the number of edges between two contiguous groups of nodes on the boundary of the circle. Another task is to determine
if there is a link connecting groups of nodes.
The edge bundled drawings are certainly task-faithful for these tasks. However, it is not information faithful, as the original graph is no longer reconstructable from the bundled layout.

\subsection{Change faithfulness}
The intuition behind \emph{change faithfulness} is that a change in
the visual representation should be consistent with the change in the
original data. Note that this is a different concept to the mental
map~\cite{peter1991mentalmap} or stability\cite{paulisch1990edge}; while these
concepts are concerned with the user's interpretation of change, the
concept of change faithfulness is concerned with the geometry of
change.

Change faithfulness is important in dynamic settings, such as interactive or streamed graph drawing. However, it is also valid in static settings, because the difference between two pictures should be consistent with the difference between the two data items that they represent.

Consider, for example, a function $V_{\groups}$ that visualizes the interaction networks $d$ that occur between European Science in Society researchers in Health\footnote{Available at \url{http://wiki.gephi.org/index.php/Datasets}. Data use in this example are filtered for Health researchers only.}.

\begin{figure}\centering
\includegraphics[width=.46\textwidth]{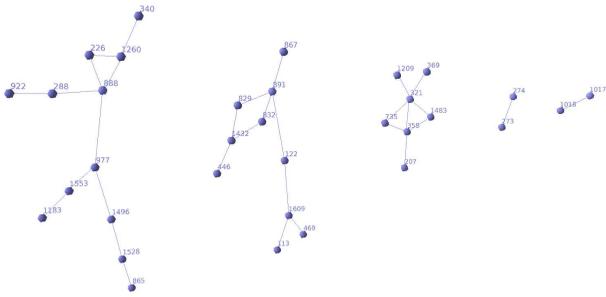}
\caption{Interaction groups between Health researchers in the EuroSiS dataset}\label{fig:eurosis}
\end{figure}

Suppose that $V_{\groups}$ uses a force directed algorithm to draw the connected components of a graph $d \in D$ separately, and arranges these components horizontally across the screen, as in Figure~\ref{fig:eurosis}. Note that $V_{\groups}$ is information faithful. However, $V_{\groups}$ is not change faithful, because a small change in the graph $d$ (such as adding an edge) can result in a large change in the picture.

\subsection{Remarks}

\subsubsection{Faithfulness and readability}
More importantly, faithfulness is a different concept to readability.
Readability is well studied in the Graph Drawing literature. It refers to the perceptual and cognitive interpretation of the picture by the viewer. Readability depends on how the graphical elements are organized and positioned. It does not depend on whether the picture is a faithful representation of the data. We can divide the readability concept into three subconcepts in the same way as we divided faithfulness:
\begin{enumerate}
  \item \emph{Information readability}: A drawing is \emph{information-readable} if the perception function $P$ is one-one; that is, if two pictures appear the same to the user, then they are pictures of the same graph. Effectively, this is saying that \emph{all} the information from the picture can be perceived by the human.
  \item \emph{Task readability}: A visualization is \emph{task-readable} for a task $T = (T_D, T_L, T_K)$ if $T_K( P (\ell) ) = T_L( \ell)$ for every layout $\ell \in L$.
  \item \emph{Change readability} is the classical \emph{mental map}.
\end{enumerate}
A good graph visualization method should achieve both faithfulness and readability; in practice, however, there may be a trade-off between the two ideals. This is especially true with large graphs, when the data size is too large for the visualization to be information-faithful; indeed, the number of pixels may be smaller than the graph size. In such cases, faithfulness is sometimes be sacrificed for readability. In specific domains, there are important tasks for which the visualization can be both readable and task-faithful.

\subsubsection{Faithfulness and determinism} 
Faithfulness is a different concept than the idea of \emph{determinism}. Determinism refers to the requirement that applying the same visualization on the same graph should give the same or similar results. For example, tree drawing algorithms are \emph{deterministic}, whereas other graph drawing algorithms, such as spring-embedder, are \emph{non-deterministic}. 

Determinism is an important criterion for visual exploration and navigation of graphs. However, a visualization method may either be deterministic or non-deterministic, while aiming for faithfulness.

\section{Faithfulness metrics \label{sec:vizmetrics}}

Faithfulness is not a boolean concept; a visualization method may be a little bit faithful, but less than 100\% faithful.
Often, one may not tell whether a picture is absolutely faithful or unfaithful. But one can compare if a picture is more faithful than the other picture of the same data.

The classical concept of readability of a graph drawing can be evaluated using a number of metrics. These include, for example, the number of edge crossings, the number of edge bends, and the area of a grid drawing. These readability metrics are formal enough that the problem of constructing a readable graph drawing can be stated as a number of optimization problems; thus optimization algorithms can be used. There is not really a clear winner amongst the metrics, and there is not a single readability metric. For example, edge crossing number does not equal to the graph readability as a graph with one more crossings is not necessarily ``less readable''.

We aim to create a list of faithfulness metrics that play the same role. Yet we should emphasize that the faithfulness metrics proposed in this section are just the examples to show the ``possibility''; they are not the \emph{only} ways to define measurement of the faithfulness concepts. Furthermore, we do not aim to create a generic ``faithfulness number'', but we try to show how quantitative measurements of the faithfulness concepts are possible.

In this section we develop a framework for such metrics.
We assume that the spaces $D$, $L$ and $R$ each has a \emph{norm}, which we generically denote by $\| \cdot \|$.
\begin{itemize}
\item 
First, each data item $d$ in data space $D$ can be modelled in multiple dimensional space $A_1 \times A_2 \times \dots \times A_i$, where $A_i$ is the $i$-th attribute dimension. A norm in data space $D$ can be Euclidean norm ($\| \cdot \|_2)$, or Manhattan norm ($\| \cdot \|_1$), or $p$-norm ($\| \cdot \|_p$). 
\item
Second, norms in result space $R$ can be defined in a similar manner of the norm in data space $D$. 
\item
Third, for layout space $L$, a definition of norms can be a bit more complex. Layout space $L$ can be modelled as $\mathbb{R}^3 \times C_1 \times \dots \times C_i$, where $C_i$ is $i$-th dimension of visual attributes. In a visualization, visual attributes of nodes include location, color, shape, transparency, texture, etc; while visual attributes of edges may further include edge bends.
\end{itemize}

Further, we assume that each of these spaces has a distance function $\Delta$ that
assigns a positive real number $\Delta(a, b)$ to each pair $a$, $b$ of
elements of the space.

\subsection{An example of information faithfulness metrics}
The simplest way to measure the information faithfulness of a graph visualization function $V$ with a specification $s$, is to measure its ``ambiguity''. 

For each data $d$ and visualization $\ell= V(d,s) \in L$, let $V^{-1} (\ell)$ denote the set $\{ (d,s) \in D \times S : V(d,s) = \ell \}$. Let $| V^{-1}(\ell) |$ denote the number of elements in $V^{-1} (\ell)$. Then the \emph{information faithfulness} of $V$ is a function $f_\text{info}$, defined by
\begin{equation}
  f_\text{info}(d,\ell) = \frac{1}{| V^{-1}(\ell) | }.
\end{equation}
The metric can have values ranging from 1 (very faithful) to 0 (unfaithful).

A more subtle approach is to measure the information faithfulness of a graph visualization function $V$ as the information loss in the channel. The loss of information during the visualization
process is defined as \emph{entropy} in information theory. The information loss is easier to measure than the total information content of a data set.
There are several techniques for measuring information content and information loss (for a full discussion, see~\cite{purchase2008theoretical}).

\subsection{An example of task faithfulness metrics}
We can measure task faithfulness as the difference between the result from the data $d$ and the result from the visualization $\ell=V(d,s)$. The \emph{task faithfulness} of $V$ is a function $f_\text{task}$, defined by:
\begin{equation}
\Delta_\text{task}(d,\ell) = \Delta (T_L(\ell) , T_D(d) )
\end{equation}
with respect to the task $T = (T_D, T_L, T_K)$ and specification $s \in S$.
One could define a normalized version of $f_\text{task}$ as:
\begin{equation}
f_\text{task}(d,\ell) = 1 / (\Delta_\text{task}(d,\ell) +1).
\end{equation}
The metric can have values ranging from 1 (very task-faithful) to 0 (task-unfaithful).

Our task faithfulness metric is kept simple enough; we neither take the transformation from data to task result, nor the interpretation of the users from the drawing to task result. In fact, different users may perceive a drawing differently in many practical cases.


\subsection{An example of change faithfulness metrics}
Tufte~\cite{tuftevisual} defines the ``lie-factor'' as the ratio of change in a
graphical representation to the change in the data. We can express Tufte's concept in terms of our model. Then, for a visualization $V$ with specification $s \in S$, the \emph{lie factor} for two distinct data $d, d' \in D$ with $d \neq d'$ is defined by:
\begin{equation*}
\lie(d', d, \ell', \ell) = \Delta(\ell,\ell')/ \Delta(d',d),
\end{equation*}
where $\ell=V(d,s)$ and  $\ell'=V(d',s)$.
Tufte's aim is to measure the quality of \emph{static} visualizations in terms of the lie-factor, but we can apply the same principle in a dynamic setting. In the ideal case (no ``lie''), the $\lie$ metric has the value of 1.

Intuitively, the lie factor increases as change faithfulness decreases, and so for two ``distinct'' data elements $d'$ and $d$ we can measure the \emph{change faithfulness} (normalized) as:
\begin{equation}
 f_\text{change}(d', d, \ell', \ell)= exp(-\lie (d',d,\ell',\ell))
\end{equation}
The values of this change metric can vary from 1 (very change-faithful) to 0 (change-unfaithful).

\section{Example 1: Multidimensional scaling and force directed approaches\label{sec:vizMDS}}

This section discusses the \emph{multidimensional scaling (MDS)}~\cite{brandes2007eigensolver} and \emph{force directed} approaches to Graph Drawing~\cite{eades,fr,finkel,north,hu} in terms of faithfulness.

The MDS approach to Graph Drawing works as follows. The input is a graph $G$=($N$,$E$), and a $|N|$ $\times$ $|N|$ matrix of \emph{dissimilarities} $\delta_{u,v}$. The goal is to map each node $u \in N$ to a point $p_u \in \mathbb{R}^k$  such that the
given dissimilarities $\delta_{u,v}$ are well-approximated by the distances $d_{u,v}$=$\|p_u - p_v\|$. The set of points $p_u$ forms the layout $\ell = V(G)$ of $G$. In practice, $k$ is commonly 2 or 3. In most applications, $\delta_{u,v}$ is chosen to be the graph theoretic distance between nodes $u$ and $v$.

To measure the success of an MDS function, a ``stress''~\cite{kruskal1964multidimensional} function is commonly used to compute the distortion between dissimilarities $\delta_{u,v}$ and fitted distances $d_{u,v}$. In
the simplest case, the stress of a layout $\ell \in L$ is:
\begin{equation}
\label{eq:stress}
\stress^{(\text{node})}(\ell)= \sum_{u\neq v}{ (\delta_{u,v} - d_{u,v} )^2},
\end{equation}
MDS can be seen as an optimization problem where the goal is to minimize this stress function.
\emph{Force directed algorithms} have a similar flavour, but view the problem as finding equilibrium in a system of forces.

\subsection{Information faithfulness}
 For most MDS approaches, there is a likelihood that vertices overlap in the ``optimal'' layout. In these cases, it is not information faithful, and different MDS methods would produce
the same result from different data sets.

\subsection{Task faithfulness}
The stress formula (\ref{eq:stress}) can be seen in terms of our task faithfulness framework.
Suppose that $T$ is a task that depends on the graph theoretic distance between nodes; let $R$ be the set of real-valued matrices indexed on the node set. For a graph $G$=($N$,$E$) $\in D$, let $T_D(G)$ be the matrix $[ \delta_{u,v} ]_{u,v \in N}$. Suppose that the visualization $V$ places node $u$ at location $p_u$; let $T_L (V(G))$ be the matrix $[ d_{u,v} ]_{u,v \in N}$. Then define
\begin{equation}
f_\text{task}^{(\text{node})}(G)= \max_{G \in D_n} \| T_L(V(G)) - T_D(G) \|_2 ,
\end{equation}
where $D_n$ is the set of graphs of size $n$ and $\| \cdot \|_2$ is the Frobenius norm. Clearly, minimising task faithfulness is equivalent to minimising the stress defined by (\ref{eq:stress}).

\subsection{Change faithfulness}
Further, we can evaluate the change faithfulness of an MDS method. In fact, MDS methods have been used extensively in dynamic settings, using stress to preserve the mental map. Suppose that at time $t$, we have a graph $G^{(t)}$, and the visualization function places node $u$ at point $p^{(t)}_u$ at time $t$.
A stress function can calculate the difference between the layout $\ell^{(t)}  \in L$ at time $t$ and the layout $\ell^{(t')}  \in L$ at an earlier time $t'$:
\begin{equation*}
\stress^{(\text{node})}(\ell^{(t)}, \ell^{(t')})= \sum_{u \in N}{ \| p^{(t)}_u - p^{(t')}_u \| ^2}.
\end{equation*}
In the so-called ``anchoring'' approach, $t'$ is zero; in the ``linking'' approach, $t'$ is the previous time frame before $t$ (see~\cite{brandes2012quantitative}).
These measures, however, aim for the mental map preservation - or change readability - rather than change faithfulness. For example, they aim to ensure that if the change in the graph is \emph{small}, then the change in the layout is \emph{small}. They do \emph{not} ensure that if the change in the graph is \emph{large}, then the change in the layout is \emph{large}.

However, we can use the stress approach to define the lie factor, such as:
\begin{equation*}
\lie (d^{(t')},d^{(t)},  \ell^{(t')}, \ell^{(t)})=  
 \frac{ \sum_{u\neq v}{ \| (d^{(t)}_{u,v} - d^{(t')}_{u,v} )/ d^{(t')}_{u,v} \| } } { \sum_{u\neq v}{ \| (\delta^{(t)}_{u,v} - \delta^{(t')}_{u,v} )/ \delta^{(t')}_{u,v} } \| }\label{eqn:changefaith_mds_sumoversum}\\
\end{equation*}
where $d^{(t)}_{u,v} = \| p^{(t)}_u - p^{(t)}_v \|$ and $\delta^{(t)}_{u,v}$ denotes the graph theoretic distance between $u$ and $v$ in $G^{(t)}$.

Then the normalized change faithfulness can be measured in terms of the distortion of the data change relative to the layout change:
\begin{equation*}
f_\text{change}(d^{(t')},d^{(t)},\ell^{(t)}, \ell^{(t')})= exp (-\lie (d^{(t')},d^{(t)}, \ell^{(t')}, \ell^{(t)}))
\end{equation*}

\subsection{Remarks}
The force directed and MDS approaches has had considerable impact on the commercial world, despite the fact that they do not have explicit or validated readability goals. We believe that the success of these approaches is due to the fact that they have explicit and validated task faithfulness goals with respect to tasks that depend on graph theoretic distances. We suggest that better MDS methods could be designed by optimising their change faithfulness using the lie factor stress above.

\begin{figure*}\centering
\subfloat[FDEB~\cite{holten}]{
\begin{minipage}[b]{0.4\linewidth}\centering%
{\label{fig:fdeb}\includegraphics[width=\textwidth]{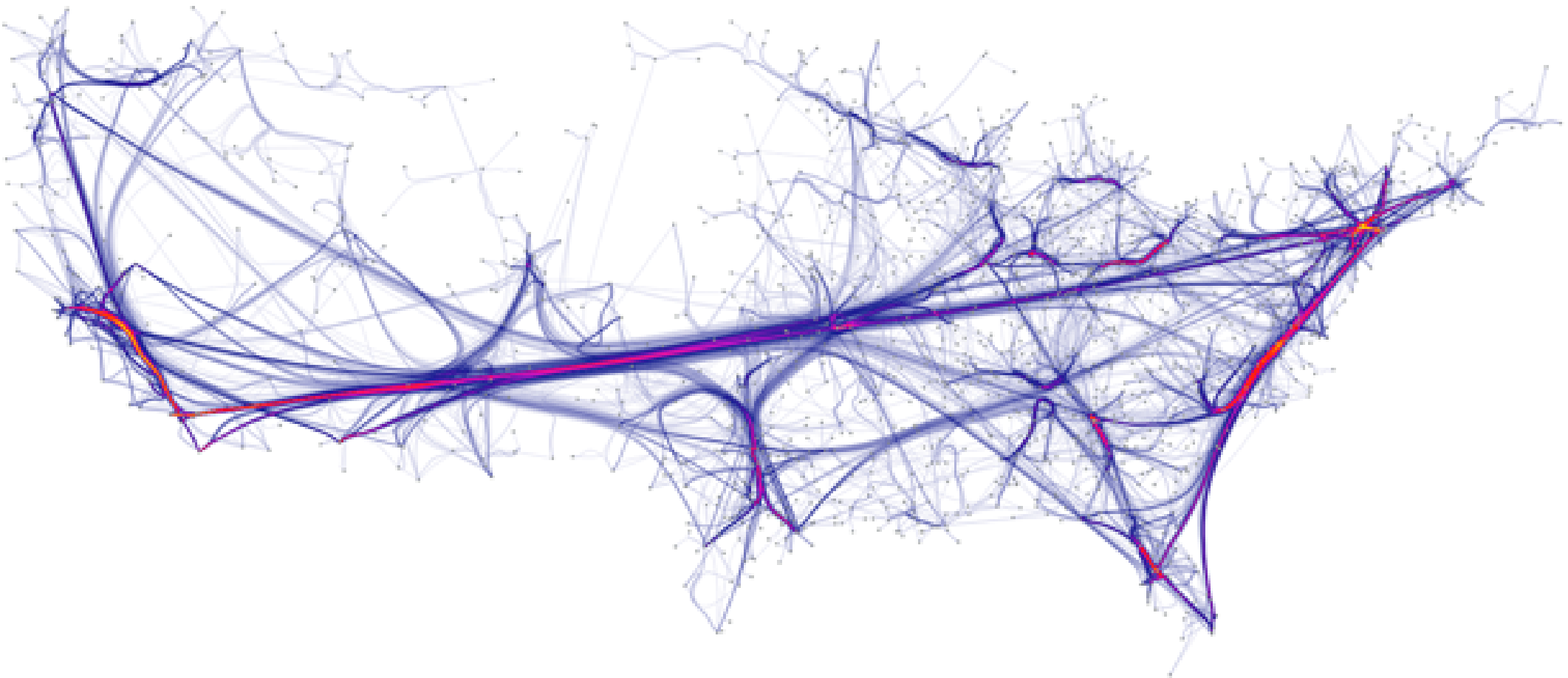}}
\end{minipage}
}\hspace{1.2cm}
\subfloat[IBEB~\cite{telea2010image}]{
\begin{minipage}[b]{0.4\linewidth}\centering%
{\label{fig:ibeb}\includegraphics[width=\textwidth]{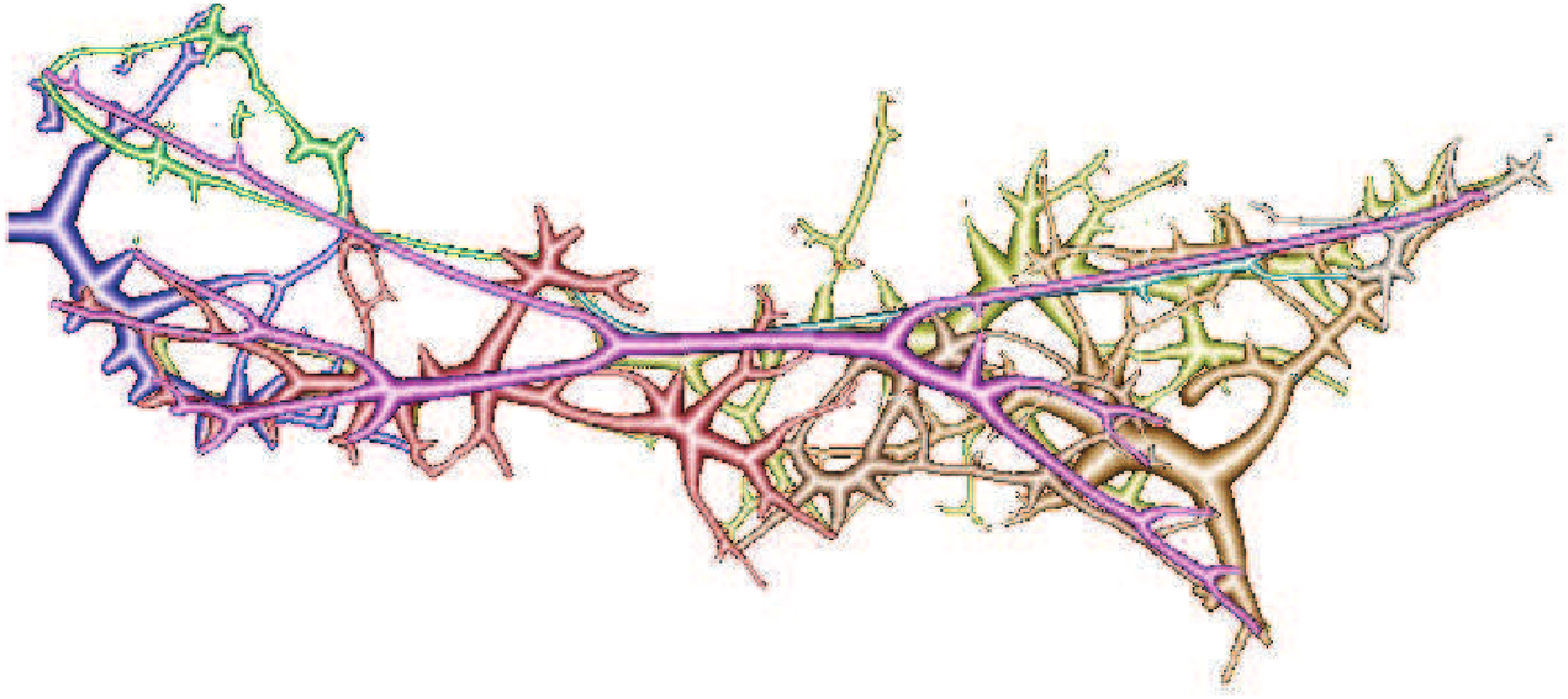}}
\end{minipage}
}\hspace{2.3cm}
\caption{Examples of US airline network visualizations using edge bundling}\label{fig:bundlingair}
\end{figure*}

\begin{figure*}\centering
\subfloat[cycle vs. stress ratio]{
\label{fig:cycle_stressratio}\includegraphics[width=.35\textwidth]{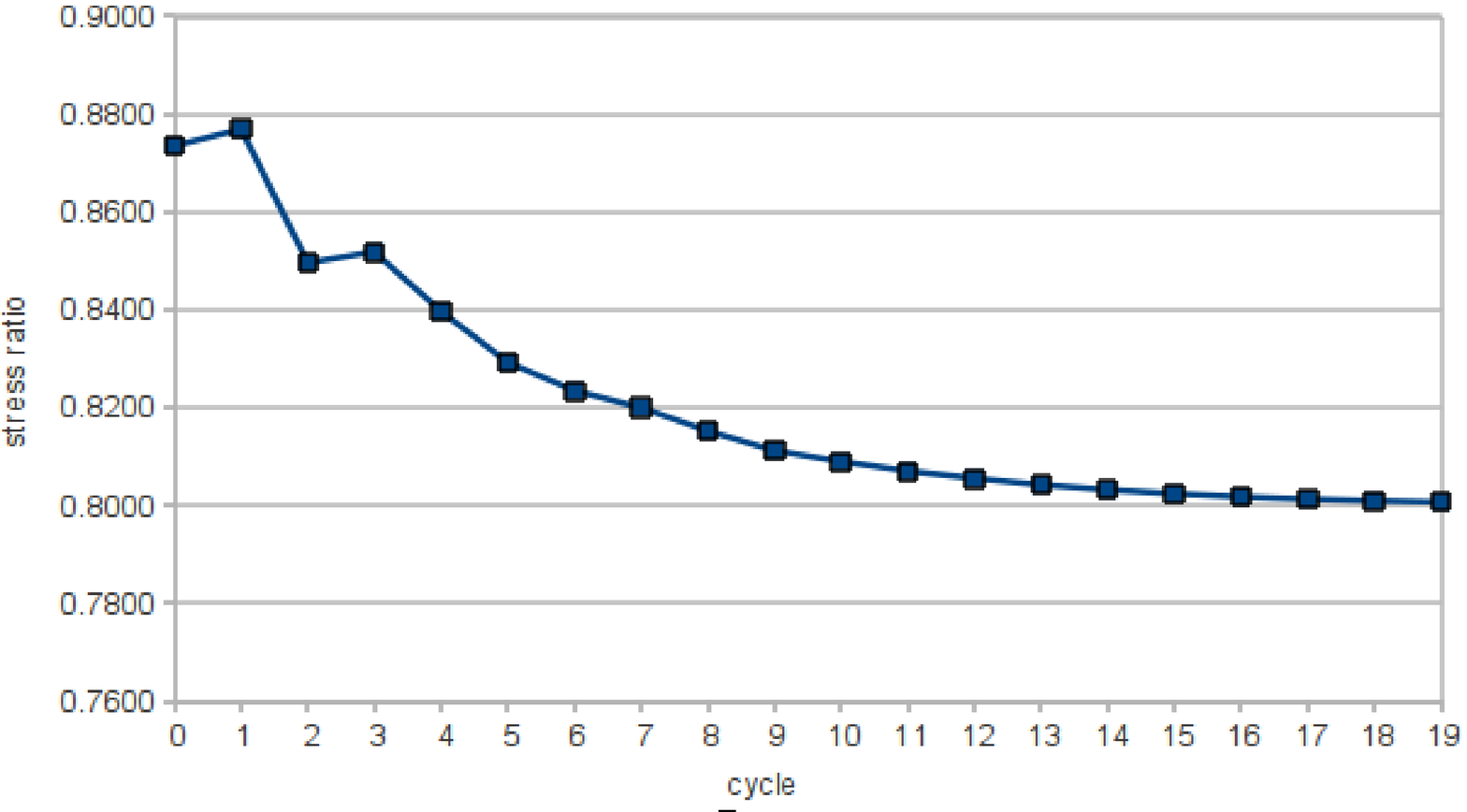}
}\hspace{1.7cm}
\subfloat[number of control points vs. stress ratio]{
\label{fig:controlpoints}\includegraphics[width=.35\textwidth]{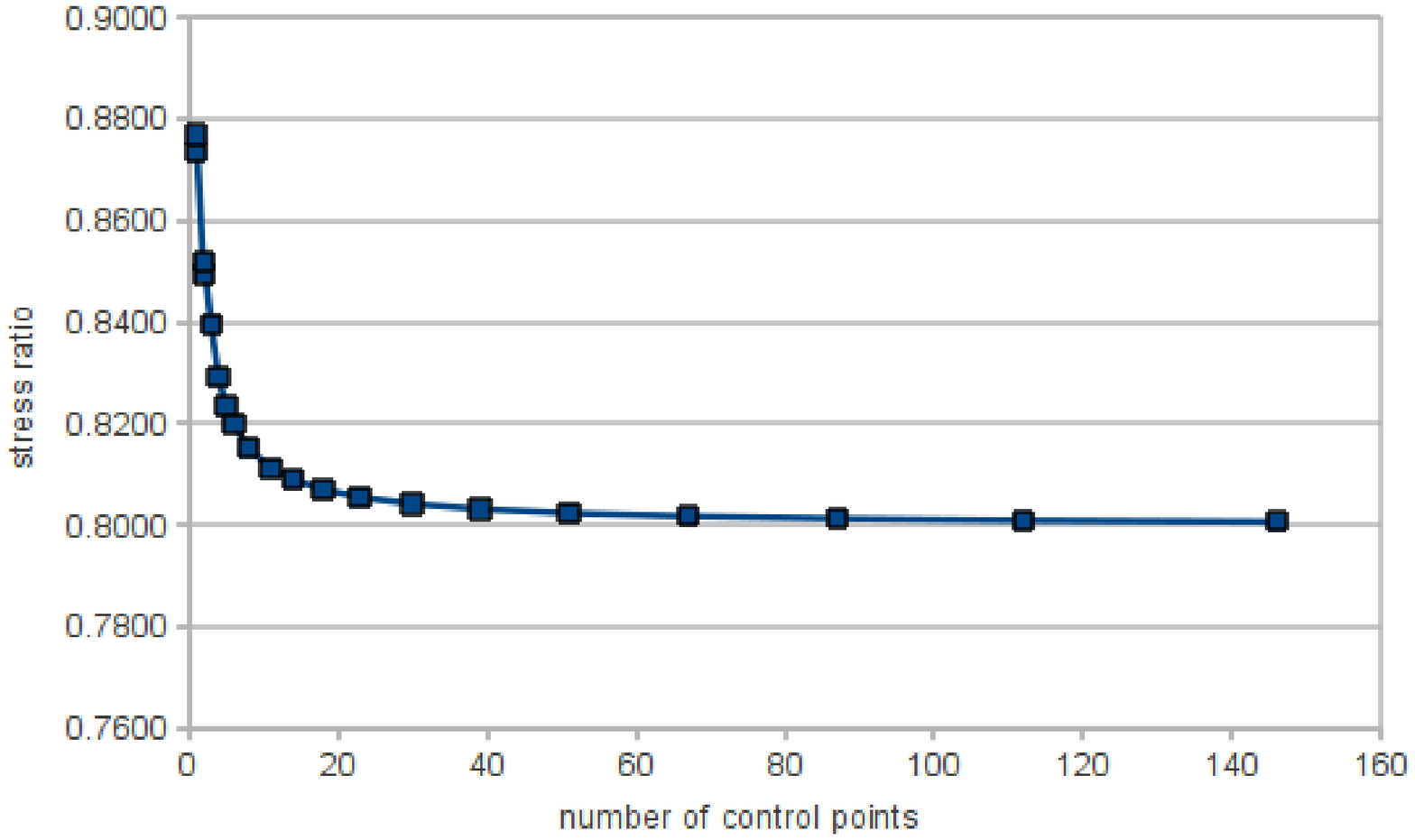}
}
\caption{Comparisons of the faithfulness of the edge bundled worldcup visualizations}\label{fig:worldcup_faithcompare}
\end{figure*}

\section{Example 2: Edge bundling\label{sec:vizbundling}}

Edge bundling, as illustrated in Figure~\ref{fig:origbundling} and \ref{fig:origbundling1}, has been extensively investigated to reduce visual clutter in graph visualizations. Many edge bundling techniques have been proposed, including \emph{hierarchical} edge bundling~\cite{holten2006hierarchical}, \emph{geometry-based} edge clustering~\cite{holten2006hierarchical,zhou2008energy,lambert2010winding}, \emph{force-directed} edge bundling~\cite{holten} and \emph{multi-level agglometive} edge bundling~\cite{gansner2011multilevel}, just to name a few.

The US airline networks are typical examples used for demonstration of edge bundling methods. Figure~\ref{fig:bundlingair} shows several edge bundling results.

Edge bundling seems to increase \emph{task readability} with respect to some tasks; for example, the classic bundling of air traffic routes in the USA (see~\cite{holten,telea2010image}) seems to make it easier for a human to identify the main hubs and flight corridors. However, some readability metrics are sacrificed; for example, the number of bends is increased, making individual paths difficult to follow (the authors are not aware of any human experiments that measure readability for edge bundling). In this Section we make some remarks about the \emph{faithfulness} of edge bundling.

\subsection{Information faithfulness}
Edge bundling reduces information faithfulness: as more edges are bundled together, it becomes harder to reconstruct the network data from a bundled layout. We can propose a rough-and-ready metric for this reduction based on the model presented in Section~\ref{sec:vizmetrics}.
Given an input graph $G = (N,E)$, an edge bundling visualization process $V$ partitions $E$ into \emph{bundles} $E = B_1 \cup B_2 \cup \ldots \cup B_k$. Let $G_i$ denote the subgraph of $G$ with edge set $B_i$ and node set $N_i$ consisting of endpoints of edges in $B_i$. Edge bundling methods ensure that $G_i$ is bipartite; suppose that $N_i = X_i \cup Y_i$ is the bipartition of $G_i$. In the bundled layout, $G_i$ is indistinguishable from a complete bipartite graph on the parts $X_i$ and $Y_i$. This representation has inherent information loss. Let $x_i = | X_i |$, and $y_i = | Y_i |$. The number of (labelled) bipartite graphs with parts $X_i$ and $Y_i$ is $2^{x_i y_i}$. Thus if $q = \sum_{i=1}^k x_i y_i$, then there are $2^q$ graphs that have the same layout as $G$. This can be used as a simple model for computing the information faithfulness of $V$.

\subsection{Task faithfulness}
Most bundling methods use a \emph{compatibility} function; roughly speaking, a compatibility function $C$ assigns a real number $C(e, e')$ to each pair $e$, $e'$ of edges. Two edges $e$ and $e'$ are more likely to be bundled together if the value of $C(e,e')$ is large. A number of compatibility functions have been proposed and tested;  these include {\em spatial} compatibility~\cite{holten} from length, position, angle and visibility between edges, {\em semantic} compatibility, {\em connectivity} compatibility, {\em importance} compatibility and {\em topology} compatibility. Some of these functions depend only on the input graph $G$, and some depend also on the layout of $G$.

For a number of tasks, such as identifying hubs in a network, highly compatible edges are equivalent; the correct performance of such tasks does not depend of distinguishing between them. Here we show that stress functions can be used to define metrics for computing the task faithfulness of the edge bundled layout relative to such tasks.

Given a pair of edges $e$ and $e'$ in an input graph $G$, let $C(e,e')$ denote their compatibility. We assume that this compatibility function depends only on $G$ and not on its layout. Let $\ell \in L$ be the layout of $G$. For two edges $e$ and $e'$, let $d(e,e')$ be the distance between the curves representing $e$ and $e'$ in $\ell$. We can choose from a number of distance functions for curves, such as the \emph{Fr\'{e}chet distance} and several distance measures for point sets.
The stress in $\ell$ is then defined as:
\begin{align}
\stress^{(\text{edge})}(\ell)= \sum_{e\neq e'}{ (C(e,e') - d(e,e'))^2}.
\end{align}

\subsection{Remarks}
Despite the plethora of recent papers in edge bundling~\cite{hurtersmooth}, there are few evaluations of effectiveness. Using the formal models outlined above, one can begin to evaluate faithfulness and compare bundling methods. For example, one can test the following hypotheses:
\begin{Hypothesis}
The force-directed edge bundling (FDEB)~\cite{holten} methods and its force-directed variants~\cite{kienreich2010application,nguyen2011tgi,selassie2011divided} are task-faithful.
\end{Hypothesis}
\begin{Hypothesis}
Force-directed edge bundling methods are {\em more} task-faithful when using {\em more} control points per edge.
\end{Hypothesis}

We have conducted several experiments with the faithfulness metrics. We use the worldcup data\footnote{Data available at http://gd2006.org/contest/WCData/} as our examples.
Our initial studies using the metrics above have shown a confirmation of the above hypotheses.
Figure~\ref{fig:worldcup_faithcompare} shows statistics of stress values at different iteration cycles of force-directed edge bundling algorithm (FDEB)~\cite{holten}. The figure has shown that FDEB achieves more faithful results when more number of iterations are performed. Furthermore, in later iterations, the result becomes more faithful as the algorithm uses more control points to improve its layout.


\section{Example 3: Visualization metaphors\label{sec:vizmetaphors}}
This section discusses our new notions of faithfulness for several representative graph visualization metaphors.

\subsection{Matrix representation}
Besides node-link diagrams, visualization of graphs as matrices form is also popular. In fact, matrix and node-link diagrams have different characteristics and can be used as suitable representations for different tasks and datasets.

Generally speaking, matrix metaphor is very faithful. All the nodes and edges are represented in the visualization. Matrices do not suffer from node overlapping and edge crossing.

However, matrix metaphor is not very task-readable for several tasks. 
Ghoniem et al.~\cite{ghoniem2005readability} report their studies on the performance of matrix and node-linked diagrams for several low level tasks. Their results show that node-link diagrams are in favour of very small (20 vertices or less) and sparse networks. Ghoniem et al. also show that matrices are more effective even for large graphs, except when the tasks are related to path tracing. Path-related tasks (such as path tracing between nodes or finding a shorted path for a pair of nodes) are the weakness of matrices; this known problem is reported  in the study of Ghoniem et al.~\cite{ghoniem2005readability}.

For certain tasks, matrix representations are more task-readable than node-link diagrams. For example, for tasks such as locating and selecting nodes, this representation is more appropriate as node labels are often more readable. Matrices do not suffer from edge crossing, which is the most trouble-some for viewing node-link visualization of dense graphs. Lastly, matrices could give an immediate overview of the sparse and dense regions within a network as well as the directedness of the connections. 

A number of methods aim to improve readability of the matrix metaphor. For example, reordering columns/rows to show highly connected groups of nodes~\cite{peng2004clutter,henry2006matexplorer} increases readability without sacrifice faithfulness.

\begin{figure}\centering
\includegraphics[width=.4\textwidth]{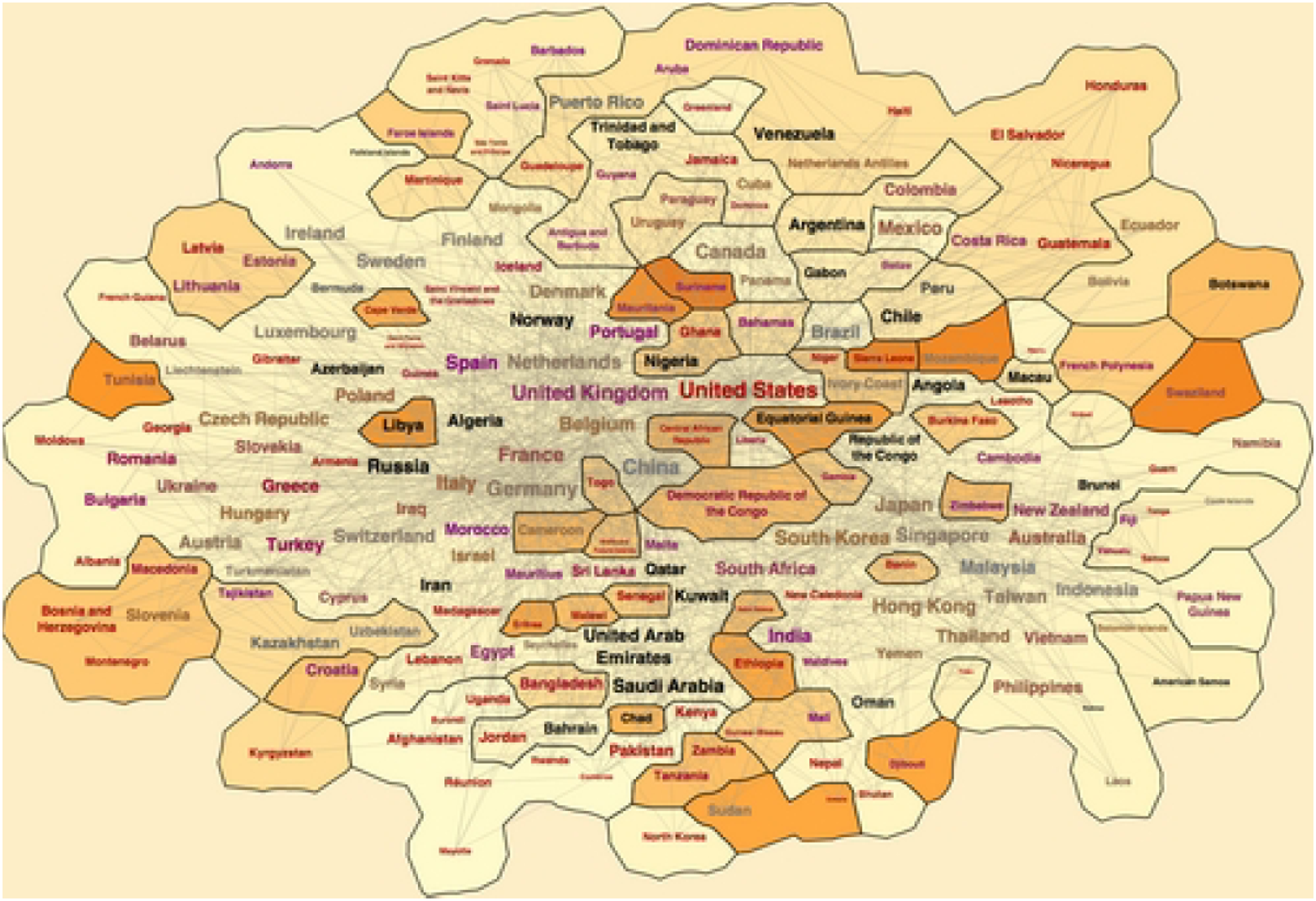}
\caption{Trade map between countries~\cite{gansner2010gmap}}\label{fig:gmap}
\end{figure}

The matrix metaphor often requires a quadratic space (of the number of nodes) to display. To visualize large graphs, information-reduction methods such as collapsing rows and columns are applied. However, such information-reduction methods increase readability and sacrifice faithfulness.

\subsection{Cartography}
Map-based approach is a promising way to produce appealing visualizations of graphs~\cite{gansner2010gmap,hu2012embedding,gronemann2012drawing}. 
In map-based visualization, a set of ``countries'' are drawn on the plane (see Figure~\ref{fig:gmap}). Each country encapsulates a node or a set of nodes within its boundary. Edges are then drawn to connect its adjacent nodes.
The visualization produced by this approach appears appealing and looks like a typical geographic map that ones usually see. Maps have been studied to display changes and trends in dynamic data~\cite{mashima2012visualizing} and stream data~\cite{gansner2012visualizing}.

\begin{figure}\centering
\includegraphics[width=.5\textwidth]{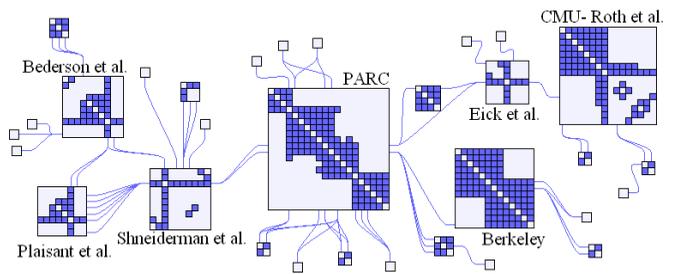}
\caption{A Hybrid Visualization of Social Networks~\cite{henry2007nodetrix}}\label{fig:nodetrix1}
\end{figure}

These map-based approaches increase task faithfulness for many tasks, such as identifying clusters or similar topics. However, map-based approaches sometimes sacrifice information faithfulness. For example, links with small weights are sometimes discarded to enable users to focus on more important links and to create appealing maps with more readable boundaries.

\subsection{Compound visualizations\label{sec:compound}}
Compound visualization techniques combine several types of visualization metaphors into the final result. Examples of compound visualizations include MatLink~\cite{henry2007matlink} and NodeTrix~\cite{henry2007nodetrix}, which integrate matrix views with node-link diagrams.

Figure~\ref{fig:nodetrix1} shows an example of the hybrid visualizations by NodeTrix~\cite{henry2007nodetrix}. 
NodeTrix proposed by Henry et al.~\cite{henry2007nodetrix} integrates node-link diagrams to show the global structure of the network and matrix-representation of groups of nodes. Thus, the method reduces the complexity and clutter of the network, while still providing all the information. 

In general, these techniques~\cite{henry2007matlink,henry2007nodetrix} seem to increase readability for some parts of the networks that are of most interest and at the same time may sacrifice faithfulness and readability in other parts of the network that are less important.
Furthermore, there can be a trade-off between global readability and local readability, and between global faithfulness and local faithfulness.

\section{Discussions and future work}
This section discusses some important factors in visualization that have direct impacts on visualization research, and their implications on the faithfulness concepts.

Research in visualization in general and graph visualization in particular is influenced by two main factors: 
(1) display capacity (hardware), and 
(2) interaction.
The former is related to the availability of display resources.
The latter refers to the development of novel interaction methods to adapt the main display parameters such as the level of details, data selection and aggregation. These factors have a major impact on how the data is presented.
We discuss each of these below.

\subsection{Faithfulness and correctness}
We should stress that faithfulness is a different concept to the classical idea of \emph{correctness} of an algorithm. An algorithm is \emph{correct} if it does what it is required to do; correctness is an essential property of every algorithm. However, a visualization method may do exactly what it is required to do without achieving faithfulness.

\subsection{Faithfulness in space and time}
The concepts of faithfulness (information-, task- and change-faithfulness) are mostly concerned with the ``\emph{space}'' dimension. That is, the visual mapping from data $d \in D$ to image $l=V(d) \in L$ does not (explicitly) consider the ``\emph{time}'' dimension.

We can extend the faithfulness concepts to integrate the time dimension.
Let $T$ denote the time domain and let $t \in T$ be a time point.
A graph $d$ at time $t$ can be presented by a two-dimensional data item $(d,t)$ of the graph $d$ and the time $t$.
The visualization process $V$ transforms the two-dimensional data item into:
\begin{itemize}
\item a drawing $V(d,t)$ in which the drawing of $V(d)$ is placed at a location in space determined by $t$ (in small-multiple display approach).
\item a drawing $V(d)$ at a time frame $V(t)$ (in animation approach).
\end{itemize}

Dynamic graph visualizations often considers the sequential number of the graph $d$ in the sequence of input graphs as the time $t$; in other words, $V(t)=t$ is an identity function.
Thus, the faithfulness concepts may disregard the time factors in these cases without loss of accuracy.
However, when time $t$ is considered in a more general setting, we should consider the followings:

{\bf Small-multiple approaches:} (1) The Information faithfulness should consider the reversibility of the time $t$; for example, placing graph elements of $d$ at a time $t$ close together and avoid mixing elements of different times. (2) Task faithfulness should consider the accuracy of task performance regarding the time attributes. For example, one should correctly identify if two data elements belong to the same time or not. (3)
Change faithfulness should further consider the change in time in the final visualizations. For example, two graphs $d$ at time $t$ and $d'$ at time $t'$ are placed close together if $| t- t'|$ is small; or placed far if $| t -  t' |$ is large.

{\bf Animation approaches:} The transformation of time $t$ to $V(t)$ can be, for example, the identity function, a linear function, a sequence-based function, or a non-linear function. (1) Information faithfulness should consider the inversibility of the time $t$; for examples, place graph elements of $d$ at a time $t$ in a separate frame $t$. (2) Task faithfulness should consider the accuracy of task performance regarding the time attributes. For example, one should correct identify whether or not a graph element exists at time $t$. (3) Change faithfulness should further consider the change of graphs together with the change in time in the animation. For example, two graphs $d$ at time $t$ and $d'$ at time $t'$ appear at frames $V(t)$ and $V(t')$ that are close/far in the animation if $| t- t'|$ is small/large.

\begin{table*}
\centering\small\addtolength{\tabcolsep}{0pt}
\caption{Faithfulness and readability}\label{table:vizFaithSummary}
\begin{tabular}{| p{.12\textwidth}| p{.38\textwidth}| p{.38\textwidth}|}
\hline 
 & \multicolumn{1}{c|}{\bf Faithfulness} & \multicolumn{1}{c|}{\bf Readability} \\[.1cm]\hline
\multirow{4}{*}{\bf Information} 
 & 
 \begin{itemize}
    \item The data set is faithfully represented by the picture. 
    \item All the original data is in the picture.
    \item The visualization process is injective.
    \end{itemize} 
 &  \begin{itemize}
    \item The user perceives the all the data in the picture.
 	\item The perception process is injective.
    \end{itemize}
\\[-.2cm]\hline 
\multirow{2}{*}{\bf Task} 
 &  \begin{itemize}
    \item The picture contains enough data to correctly perform the task
    \end{itemize} 
 & \begin{itemize}
    \item The user perceives enough data from the picture to correctly perform the task.
    \end{itemize} 
\\[-.2cm]\hline
\multirow{3}{*}{\bf Change} 
 &  \begin{itemize}
    \item Changes in the picture are consistent with changes in the data.
    \end{itemize} 
 & \begin{itemize}
    \item The mental map is preserved.
    \item The user can remember one screen from the previous screen.
    \end{itemize} 
\\ \hline
\end{tabular}
\end{table*}

\begin{table}\centering
\caption{Faithfulness of existing visualization methods}\label{table:vizFaithVisualizationSummary}
\begin{tabular}{|l|c|c|c|c|c|c|}\hline
\cline{2-7}
  & \multicolumn{3}{|c|}{\bf Faithfulness} & \multicolumn{3}{|c|}{\bf Readability} \\[.2cm] \cline{2-7}
\bf Visualization & \rotatebox{90}{\bf Info} & \rotatebox{90}{\bf Task} & \rotatebox{90}{\bf Change} & \rotatebox{90}{\bf Info} & \rotatebox{90}{\bf Task} & \rotatebox{90}{\bf Change} \\[.2cm]
\hline
Force-directed & - & + & + & - & + & - \\
MDS & - & + & + & - & + & - \\
Edge concentration & - & + & - & - & + & - \\
Confluent drawing & - & + & - & - & + & - \\
Edge bundling & - & + & - & - & + & - \\
Matrix metaphor & + & + & + & + & -  & - \\
Map-based metaphor & - & + & + & + & + & - \\
Combination & - & + & + & - & + & + \\
\hline
\end{tabular}
\end{table}


\subsection{Display device}
Although there have been numerous revolutions in hardware technologies, screen size is still a precious but limited resource. Technically speaking, ``screen size'' refers to the number of pixels in a display rather than the physical  dimensions of the screen. Examples of displays include the high-resolution displays, large-scale power walls and small portable devices. 

Faithfulness metrics should be addressed in a specific formula depending on the characteristics of the available output devices.
For visualizing large data sets, the large number of data elements can compromise performance or overwhelm the capacity of the viewing platform.
In addition, visualization methods should take user-centric requirements (from user inputs or interactions) into account to produce best results that balance between faithfulness and readability.

\subsection{Interaction}
Ideally, the user interface and interaction techniques should be simple enough to help users to perform the tasks with less cognitive efforts; a complex user interface may sometimes distract the users.
Novel interaction techniques need to seamlessly support visual communication (user inputs) of the user with the system.
For example, when a user may need to explore and edit a certain part of the graph several times, the system can capture the users' area of interest. The layout adjustment algorithms should adapt to user needs and should cleverly sacrifice the overall faithfulness to the local faithfulness (of the part of the graph of user interests).


The below are the implications of our faithfulness concept from three specific types of interactions.

{\it Affine transformation:} Intuitively, faithfulness does not depend on global affine transformations such as rotation, translation and scaling. As long as the drawing is consistent with the data, the visualization is considered faithful. 
On the other hand, a (partial) transformation of the local parts of the graphs may increase readability of these parts of the graphs, yet may degrade faithfulness of other parts of the graphs.

{\it Distortion techniques:} such as lens effects and occlusion reduction can provide the analyst with trade-off between faithfulness and visual clarity. These methods typically transform object's position to improve local readability at the cost of accuracy of global relations.

{\it Level of detail:} Relevant data patterns and relationships can be visualized in several levels of detail at appropriate levels of data and visual abstraction. The overall goal is to balance between faithfulness and readability to maximize user expectations.

\section{Concluding Remarks\label{sec:vizconclusion} }

This paper has introduced the concept of faithfulness
for graph visualization. We believe that the classical
readability criteria are necessary but not sufficient
for quality graph drawing; faithfulness is a generic
criterion that is missing. 

The paper has described the
faithfulness concept in a semi-formal model.
We have distinguished three kinds of faithfulness: information faithfulness, task faithfulness, and change faithfulness.
Table~\ref{table:vizFaithSummary} gives a summary of these faithfulness concepts.

Based on the visualization model and the faithfulness concepts, we have also presented a model for faithfulness metrics. 
In Section~\ref{sec:vizMDS},~\ref{sec:vizbundling} and~\ref{sec:vizmetaphors}, we
illustrate the faithfulness concept with three examples.
\begin{itemize}
\item The first example is multidimensional scaling / force-directed methods. We believe that future directions
of these methods would need to balance the aims of
readable outputs versus faithful representations. 

\item The
second example is edge bundling. Despite a recent
upsurge of interest in edge bundling, there are very
few evaluations; we show that faithfulness metrics
may prove the key to evaluation. 

\item The last example
includes matrix metaphors and map-based visualizations. These methods are useful for large graphs; we
show that future directions of these methods would
need to balance between global / local readability
versus global / local faithfulness.
\end{itemize}

\subsection{Guidelines}
Table~\ref{table:vizFaithVisualizationSummary} summarizes the faithfulness of some selected visualization methods. In this table, `$-$' represents `no' or `low'; `$+$' represents `yes' or `high'. These results may give readers a reference, which may not be the same with the readers' viewpoints. 

A `$+$' for task faithfulness should be interpreted as: a method is faithful for some tasks, rather than faithful for all tasks. In Section~\ref{sec:vizMDS},~\ref{sec:vizbundling} and~\ref{sec:vizmetaphors}, we describe several specific tasks, in which these visualizations are task-faithful.

\subsection{Remarks on 3D drawings}
We conclude this paper with a remark about 3D graph drawing. The occlusion problem for 3D means that, even
with binocular displays, some part of the graph is always hidden. This can be seen as a lack of not only readability but also information faithfulness. We suggest that the lack of commercial impact of 3D graph drawing is partially due to its inherent lack of faithfulness.

\bibliographystyle{abbrv}
\bibliography{faith}

\vspace{-1cm}


\begin{IEEEbiographynophoto}{Quan Hoang Nguyen}
Quan Nguyen is a post-doc research fellow at the School of Information Technologies, University of Sydney, Australia.
\end{IEEEbiographynophoto}
\begin{IEEEbiographynophoto}{Peter Eades}
Peter Eades is an Emeritus Professor at the School of Information Technologies, University of Sydney, Australia.
\end{IEEEbiographynophoto}
\vfill

\clearpage
\enlargethispage{-5in}

\appendices

\end{document}